\documentclass[lettersize,journal]{IEEEtran}

\usepackage{amsmath,amsfonts}
\usepackage{algorithm}
\usepackage{algpseudocode}
\usepackage{graphicx}
\usepackage{booktabs}
\usepackage{array}
\usepackage[table]{xcolor}
\usepackage{enumitem}
\usepackage{pifont}
\usepackage{tikz}
\usepackage{etoolbox}

\usepackage{cite}
\usepackage[colorlinks=true,linkcolor=blue,citecolor=blue,urlcolor=blue,filecolor=blue]{hyperref}

\setlist[itemize]{leftmargin=1em,labelsep=0.5em,noitemsep,topsep=0pt}
\setlist[enumerate]{leftmargin=1em,labelsep=0.5em,noitemsep,topsep=0pt}

\definecolor{emptycirc}{RGB}{255,255,255}
\definecolor{halfcirc}{RGB}{0,0,0}
\definecolor{fullcirc}{RGB}{169,169,169}

\newcommand{\customcirc}[1]{%
  \tikz[baseline=(char.base)]{%
    \node[shape=circle,draw,inner sep=2pt,fill=#1] (char) {};%
  }%
}

\begin{document}

\title{GenTI: Benchmarking LLMs for Autonomous IDPS Rule Generation for Unseen Attacks}

\author{Hassan Jalil Hadi,  Rehana Yasmin, Ali Shoker
\thanks{Hassan Jalil Hadi, Rehana Yasmin, and Ali Shoker are with the Cyber Security and Resilience Technology (CyberSaR), King Abdullah University of Science and Technology (KAUST), Saudi Arabia (e-mail: hassan.hadi@kaust.edu.sa, alishoker@kaust.edu.sa).}

}

\maketitle
\begin{abstract}
Rule-based Intrusion Detection and Prevention Systems (IDPS) offer precise attack detection as well as mitigation, however their manually crafted, signature-driven rules limit adaptability to emerging and zero-day threats. Additionally, existing public datasets (e.g., CICIDS2017, UNSW-NB15) focus on traffic classification and provide little structured information to support automatic rule synthesis or prevention logic. To address this gap, we propose Generative Thread Intelligence (GenTI) \footnote{GenTI refers to the proposed framework, and GTI refers to the dataset.} an LLM-driven benchmark for automatic generation of IDPS rules targeting  unseen attacks. The dataset (GTI) aggregates over 150k detection and prevention rules from Snort, Suricata, Emerging Threats, as well as 50k YARA, each annotated with protocol behavior, payload signatures, contextual relationships, mappings to Cyber Threat Intelligence (CTI), along with actionable response types (alert, drop, reject). Moreover, on top of this corpus we design an LLM-based pipeline that transforms analyst prompts and representative payloads into deployable rules via structured prompt engineering, Chain-of-Thought (CoT) reasoning, as well as a Chain-of-Verification (CoVe) loop for syntactic, semantic, and security validation. The generated rules are executed in real time on (Snort/Suricata) and evaluated by syntax accuracy, semantic similarity, CTI coverage, security effectiveness as well as unseen attacks detection. Furthermore, our GenTI instantiation achieves a composite rule-quality score of 89.4\%, with 94.8\% CTI coverage, improving unseen attacks detection from 45\% to 87.4\% and reducing the false-positive rate from 8.5\% to 2.3\%. Overall, GenTI establishes the first large-scale benchmark that tightly couples rule-level CTI with LLM-based automation, enabling adaptive, self-evolving IDPS.
\end{abstract}

\begin{IEEEkeywords}
 IDPS, Large Language Models (LLMs), CTI, Cybersecurity Automation, Unseen Attacks
\end{IEEEkeywords}

\section{Introduction} \label{sec:intro}
\begin{figure}[htbp]
    \centering
    \includegraphics[scale=0.22]{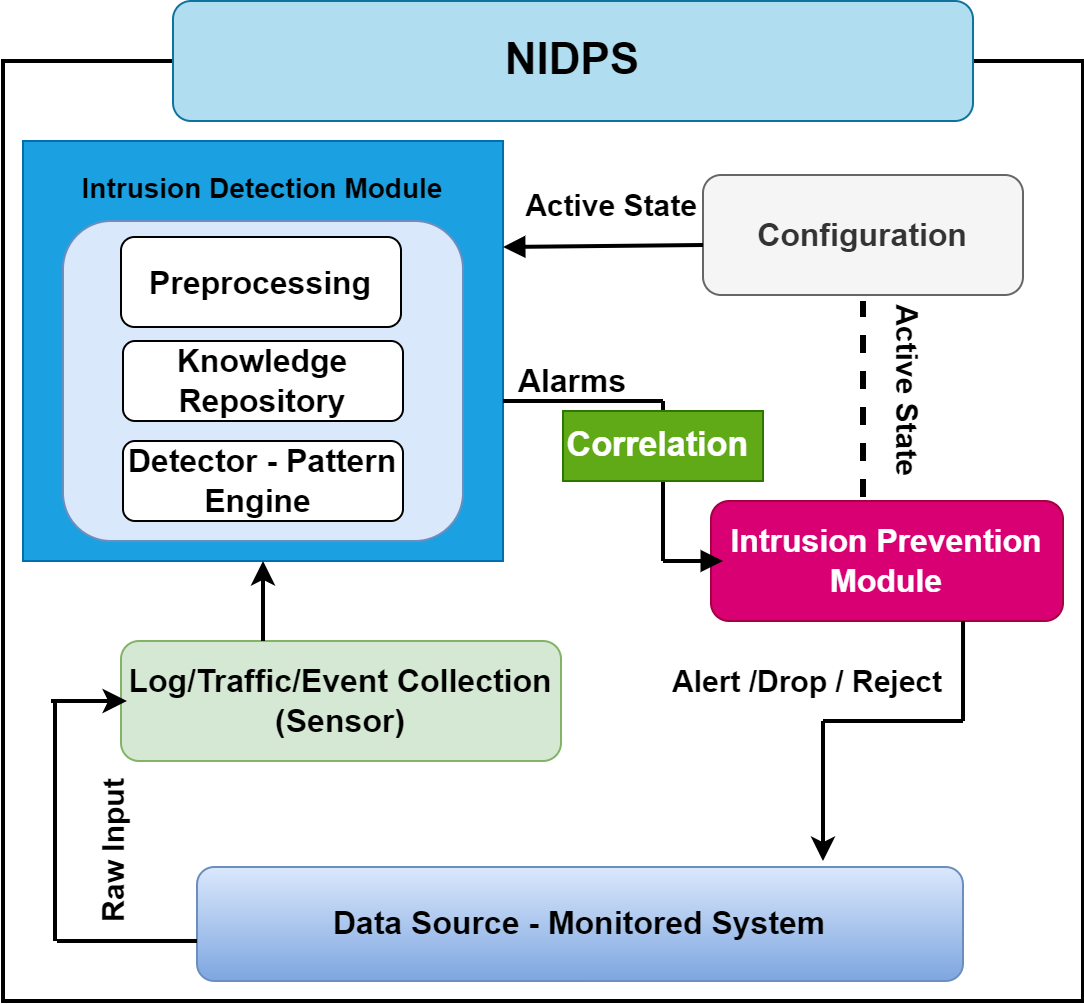}
    \caption{Core Architecture of a Basic NIDPS}
    \label{fig:01}
\end{figure}

\IEEEPARstart{I}{n} recent years, the scale and sophistication of cyber threats have continued to grow, making timely as well as accurate detection of attacks a central challenge in cybersecurity.  To mitigate these risks, organizations increasingly rely on a broad ecosystem of security solutions, including next-generation firewalls, Network Intrusion Detection Systems (NIDS), Network Intrusion Prevention Systems (NIPS) \cite{wei2023xnids}, Security Operations Centers (SOC) \cite{ofte2023understanding}, Security Information and Event Management (SIEM), endpoint detection \cite{winkler2025proactive}, EDR, XDR and response tools \cite{wei2023xnids}. Among these, NIDPS are particularly critical, as they monitor network traffic, identify suspicious or abnormal behavior, and generate alerts or protective actions to support security analysts. Also, modern NIDPS techniques can broadly be categorized into rule-based, statistical-based, behaviour-based, machine learning-based, and anomaly detection approaches \cite{he2023adversarial}, \cite{al2024analysis}. Rule-based (signature-based) systems identify threats by matching observed traffic against predefined attack patterns, triggering alerts when a rule condition is satisfied, as illustrated in Fig. \ref{fig:01}. Behavioural and statistical methods typically rely on thresholds or aggregated metrics to flag deviations. Further,  anomaly detection methods learn normal traffic patterns and detect malicious activity as significant departures from this learned baseline.

Although, machine-learning-based NIDPS have demonstrated strong detection capability \cite{wei2023xnids}, they are still rarely trusted in production environments due to opaque decision processes and the large number of false alerts they may
generate \cite{al2024analysis}, \cite{aldweesh2020deep}. As a result, rule-based NIDPS remain the dominant choice in many organizations. Open-source communities provide and maintain extensive rule sets for engines such as Snort, Suricata, YARA as well as Emerging Threats, which are continuously updated and widely deployed. However, these rule sets primarily detect known attacks and offer limited protection against zero-day or previously unseen intrusion attempts. This limitation arises because rules are typically crafted manually by security experts based on already disclosed vulnerabilities and documented attack patterns.

\par{
Furthermore, attackers increasingly exploit undisclosed vulnerabilities and employ obfuscation techniques to disguise payloads and evade existing signatures. These strategies create new attack variants for which no rules currently exist. To address this challenge, numerous machine-learning-based intrusion detection approaches have been proposed to identify previously unseen attacks \cite{al2025multi}. Nevertheless, these models are usually trained on a narrow set of attack patterns from a number of public datasets, such as CICIDS2017 \cite{sharafaldin2018toward} and CSE-CIC-IDS2018 \cite{sharafaldin2018toward}, which restricts their coverage along with generalization in realistic environments. To bridge this gap, our work introduces a new benchmark dataset Generative Thread Intelligence (GTI) , designed to support automatic generation, updating, and verification of detection as well as prevention rules for IDS and IPS.
\par{
In addition, automatic generation of high-quality detection rules faces three key challenges. First, there is no large-scale, publicly available benchmark dataset tailored for IDPS research. Existing resources do not support fine-tuning LLMs on rule synthesis, refinement, or contextual reasoning. Second, most automatic rule-generation methods are designed for firewalls or header-level pattern matching. As a result, they fail to capture payload-level semantics and the detailed behaviours, tactics, and techniques used in modern attacks. Third, there is no systematic mechanism to verify the syntactic correctness and semantic soundness of automatically generated rules. It also remains unclear whether such rules can operate with acceptable false-positive rates in realistic production environments.
}

\par{
To address these issues, we propose GenTI, a benchmark dataset (GTI) and LLM-driven workflow for automatic NIDPS rule generation. GenTI  provides a large corpus of payload-aware detection as well as prevention rules, enriched with mappings to CTI sources such as OTX, MISP, and MITRE ATT\&CK. The workflow consists of seven components including, preprocessing, prompt analysis, attack–sample mapping, rule refinement and validation, CoT–CoVe reasoning, and output generation. Attack payloads and metadata are first normalized then converted into task-specific prompts. After that, a three-stage prompt design is applied in which LLMs generate the initial rules and Chain-of-Thought (CoT) prompting refines the rule logic. Then, a Chain of Verification (CoVe) integrates refinement and optimization to remove redundant rules, prune irrelevant options, and ensure syntactic along with semantic consistency, thereby reducing false positives. Building on this pipeline, the main contributions of this work are as follows: }
\begin{itemize}
    \item Firstly, to the best of our knowledge, we introduce GTI, the first benchmark dataset and CTI-enriched corpus tailored to LLM-based NIDPS rule generation. It consolidates over 150k detection as well as prevention rules from Snort\footnote{\url{https://www.snort.org/downloads}}, 
Suricata\footnote{\url{https://suricata.io/download/}}, 
Emerging Threats\footnote{\url{https://rules.emergingthreats.net/}} \footnote{\url{https://rulezet.org/}}, and 50k YARA into a unified format. It leverages LLMs to map rules to CTI sources (e.g., MITRE ATT\&CK, MISP, OTX) and to annotate Rule–Context and Relationships Annotation (RCRA), thereby providing deeper insight into attacks and their corresponding detection rules.
\item  Secondly, we incorporate YARA-style payload-level signatures and translate them into IDPS-compatible logic, enabling the automatic generation of both detection as well as prevention rules (e.g., alert, drop, reject) at the payload level. 
    \item Thirdly, we develop a four-stage progressive training pipeline comprising: (1) syntactic grounding through rule structure and header pattern learning, (2) semantic enrichment via CTI-augmented CoT reasoning, (3) CoVe-driven verification for false-positive reduction, and (4) full-complexity integration with transfer learning across stages.
    
    \item Lastly, a real-time evaluation of GenTI on open-source NIDPS engines (Snort and Suricata) is performed using realistic malicious and benign traffic, showing strong malicious-traffic detection performance across standard IDS metrics.
\end{itemize}

\section{Related work} \label{sec:re}

This section reviews related work on three key aspects including, 
(1) Existing IDS datasets, (2) Automatic rule generation for malicious traffic detection, and (3) The need for benchmark datasets that support LLM-based rule synthesis and evaluation. We also underline the limitations of current approaches that motivate the design of the proposed GenTI benchmark.
\begin{table*}[ht]
\scriptsize
\centering
\caption{\centering Comparison of Current Intrusion Detection Datasets with the GTI Benchmark. \\
\protect\customcirc{emptycirc} = ``Not Available'',
\protect\customcirc{halfcirc} = ``Partially Available'',
\protect\customcirc{fullcirc} = ``Available''.}
\label{tab:1}
\renewcommand{\arraystretch}{1.0}

\begin{tabular}{%
    >{\raggedright\arraybackslash}p{1.80cm}  
    >{\centering\arraybackslash}p{1.5cm}    
    >{\centering\arraybackslash}p{1.0cm}    
    >{\centering\arraybackslash}p{1.4cm}    
    >{\raggedright\arraybackslash}p{2.5cm}  
    >{\centering\arraybackslash}p{1.5cm}    
    >{\centering\arraybackslash}p{1.5cm}    
    >{\centering\arraybackslash}p{1.5cm}    
    >{\centering\arraybackslash}p{1.5cm}    
}
\toprule
\textbf{Dataset} &
\textbf{Rule Availability} &
\textbf{CTI Mapping} &
\textbf{RCRA} &
\textbf{Type of Attacks} &
\textbf{Rules Verification} &
\textbf{Actions Types} &
\textbf{LLM Compatibility} &
\textbf{Unseen Attacks Utility} \\
\midrule

KDD-Cup99 \cite{kddcup1999} &
\customcirc{emptycirc} & \customcirc{emptycirc} & \customcirc{emptycirc} &
DoS, Probe, R2L, U2R  &
\customcirc{emptycirc} & \customcirc{emptycirc} &
\customcirc{emptycirc} & \customcirc{emptycirc} \\

NSL-KDD \cite{nslkdd} &
\customcirc{emptycirc} & \customcirc{emptycirc} & \customcirc{emptycirc} &
DoS, Probe, R2L, U2R  &
\customcirc{emptycirc} & \customcirc{emptycirc} &
\customcirc{emptycirc} & \customcirc{emptycirc} \\

DEF-CON  \cite{defconctfarchive} &
\customcirc{emptycirc} & \customcirc{emptycirc} & \customcirc{emptycirc} &
Buffer Overflow  &
\customcirc{emptycirc} & \customcirc{emptycirc} &
\customcirc{emptycirc} & \customcirc{emptycirc} \\

CICIDS2017 \cite{sharafaldin2018toward} &
\customcirc{emptycirc} & \customcirc{emptycirc} & \customcirc{emptycirc} &
DDoS  &
\customcirc{emptycirc} & \customcirc{halfcirc} &
\customcirc{emptycirc} & \customcirc{emptycirc} \\

CIC-IDS2018  \cite{sharafaldin2018toward} &
\customcirc{emptycirc} & \customcirc{emptycirc} & \customcirc{emptycirc} &
Port Scanning &
\customcirc{emptycirc} & \customcirc{halfcirc} &
\customcirc{emptycirc} & \customcirc{emptycirc} \\

5G-NIDS \cite{samarakoon20225g} &
\customcirc{emptycirc} & \customcirc{emptycirc} & \customcirc{emptycirc} &
Botnet  &
\customcirc{emptycirc} & \customcirc{emptycirc} &
\customcirc{emptycirc} & \customcirc{emptycirc} \\

IoT-NI \cite{ullah2020scheme} &
\customcirc{emptycirc} & \customcirc{emptycirc} & \customcirc{emptycirc} &
Fuzzers, DoS, Exploits &
\customcirc{emptycirc} & \customcirc{emptycirc} &
\customcirc{emptycirc} & \customcirc{emptycirc} \\

IoT-23 \cite{ullah2021design} &
\customcirc{emptycirc} & \customcirc{emptycirc} & \customcirc{emptycirc} &
DDoS, Heartbleed, Infiltration, Brute-Force  &
\customcirc{emptycirc} & \customcirc{emptycirc} &
\customcirc{emptycirc} & \customcirc{emptycirc} \\

MQTT-IoT \cite{hindy2020machine} &
\customcirc{emptycirc} & \customcirc{emptycirc} & \customcirc{emptycirc} &
UDPFlood, HTTPFlood  &
\customcirc{emptycirc} & \customcirc{emptycirc} &
\customcirc{emptycirc} & \customcirc{emptycirc} \\

UAV-NIDDS \cite{hadi2025uav} &
\customcirc{emptycirc} & \customcirc{emptycirc} & \customcirc{emptycirc} &
DoS, MITM, Port Scan, Botnet  &
\customcirc{emptycirc} & \customcirc{emptycirc} &
\customcirc{emptycirc} & \customcirc{emptycirc} \\

\rowcolor{gray!12}
\textbf{GTI (2025)} &
\customcirc{fullcirc} 200K+ real production IDPS rules & \customcirc{fullcirc}  & \hspace{0.5em} \customcirc{fullcirc} \hspace{2.9em}  Chain of Thought &
All Types of Attacks &
\hspace{2em} \customcirc{fullcirc}      CoVe &  \hspace{0.8em} \customcirc{fullcirc} \newline Alert, Drop, Reject &
\customcirc{fullcirc} & \customcirc{fullcirc} \\
\bottomrule
\end{tabular}
\end{table*}
\subsection{Existing IDS Dataset}

Early research on NIDS has relied heavily on benchmark datasets such as DARPA \cite{lippmann2000evaluating}, KDD-Cup99 \cite{kddcup1999},  NSL-KDD \cite{nslkdd}, DEFCON \cite{defconctfarchive}, CAIDA \cite{caida2007ddos}. These datasets were primarily designed for flow or connection-level anomaly and attack classification, often in simulated or controlled network environments. Later efforts, including \cite{sharafaldin2018toward}, CSE-CIC-IDS2018 \cite{sharafaldin2018toward}, UNSW-NB15 \cite{moustafa2015unsw}, 5G-NIDS \cite{samarakoon20225g}, IoT-NI \cite{ullah2020scheme}, IoT-23 \cite{hindy2020machine}, MQTT-IoT-IDS \cite{hindy2020machine}, and UAV-NIDDs \cite{hadi2025uav}, provide more diverse and realistic traffic. They include multiple attack categories such as DoS, probing, brute force, botnets, and DDoS. A recent survey offers a concise overview of these datasets until 2024 and confirms their continued importance for evaluating NIDS performance \cite{al2024analysis}, \cite{hadi2025uav}.
\par{
However, as summarized in Table~\ref{tab:1}, these datasets are not designed for rule-level modeling or automatic new rule generation. Most of them do not release the actual IDS rules used in their experiments, nor do they provide structured mappings to CTI sources. Attributes that are crucial for LLM-based rule synthesis are typically absent, including rule availability, CTI mapping, RCRA, rule verification metadata, explicit action types (e.g., \texttt{alert}, \texttt{drop}, \texttt{reject}), and any indication of LLM compatibility or unseen attacks utility. Moreover, they mainly expose flow or header-level features (e.g., IP addresses, ports, and basic protocol statistics) and provide little or no labelled payload content. In practice, they support traffic or attack classification. However, they cannot be directly used to learn how production-grade IDPS rules, especially payload-level rules based on YARA-style content patterns, are written, refined, or validated.}
\par{
In contrast, GenTI is explicitly constructed as a rule-centric benchmark. It aggregates more than 200k real-world detection as well as prevention rules from Snort, Suricata, Emerging Threats, YARA, and attaches rich metadata for each rule, including CTI mappings, RCRA, rule verification status, and action types. Moreover, GenTI is explicitly designed for rule-level modelling at the payload level. It exposes rich, labelled payload content and YARA-style signatures, enabling LLMs to learn as well as synthesize new regular expressions with content patterns required for rule creation. This design aims to improve coverage of previously unseen attacks. These attacks manifest their behavior in payload semantics rather than only in header features. As a result, GenTI is suitable for evaluating intrusion detection performance. It is also uniquely suited for fine-tuning and assessing LLM-based pipelines that perform rule generation, refinement, and verification.}

\subsection{Approaches to Automatic Rule Generation}
Automatic generation of detection rules from attack samples can reduce the manual workload of security experts and improve the detection of unseen attacks and emerging threats. Early work on automatic rule generation relied on honeypots and protocol analysis without using machine learning. Honeycomb, for example, creates Snort rules by extracting common substrings from honeypot sessions and applying protocol conformance checks to avoid trivial patterns \cite{kreibich2004honeycomb}. Such systems reduce some manual effort, but the generated rules are typically narrow, tied to specific flows, and lack higher-level threat semantics.
\par{
Later approaches introduced machine-learning techniques to mine signatures from malicious traffic. LARGen used Latent Dirichlet Allocation on network flows to extract multiple content strings and then constructed IDS signatures from the resulting topics. Further, CMIRGen combined clustering, sequence similarity, and black-box model inference to derive token-based rules from malicious payloads, improving robustness over simple pattern matching \cite{zhang2020cmirgen}. Besides, the automatic NIDS Rule Generating System focused on HTTP-like malware communication and generated Snort rules from protocol-aware analysis of traffic traces \cite{kao2015automatic}. These methods generalised better than pure honeypot pattern mining, yet they still operate on limited attack corpora, focus mainly on flow or header features, and provide little semantic context about tactics, techniques, or vulnerabilities. 
\par{
More recently, LLMs have been used to automate IDS rule creation. Harnessing LLMs for automated rule generation in cyber ranges introduced a prompt-based framework that generates and refines rules for web attacks \cite{du2025harnessing}. Hex2Sign converts hexadecimal PCAP data from honeypots into Suricata signatures using LLMs, demonstrating rule synthesis directly from low-level traffic \cite{balasubramanian2024hex2sign}. Hu et al. designed an LLM-based agent that ingests vulnerability reports and existing rules to generate and generalize IDS signatures, showing improved detection on proprietary datasets \cite{hu2024llm}. These LLM-driven systems show that large models can reason about attack payloads and security context. However, existing work typically rely on small rule sets (roughly 400–700 IDS rules) and narrowly scoped, task-specific datasets (e.g., web attacks, vulnerability-specific traffic, or header-level features). Additionally, instead of designing fundamentally new, payload-level rules for unseen attack patterns, current methods largely regenerate similar instances of existing rules. As a result, they are not evaluated in production or real-time environments, do not release reusable rule corpora, and do not provide payload-level YARA-style signatures.
}

\subsection{Why creating a new benchmark dataset}
Upon careful examination of Table \ref{tab:1} and the current landscape of IDS datasets, we identify two critical limitations. First, existing datasets such as CICIDS2017 \cite{sharafaldin2018toward}, CSE-CICIDS2018 \cite{sharafaldin2018toward}, UNSW-NB15 \cite{moustafa2015unsw} and related corpora have greatly facilitated NIDS research. However, these datasets cover a relatively
fixed set of attack families (e.g., DoS, DDoS, brute-force, basic
web attacks) and are primarily labelled for traffic classification. Consequently, they provide limited support for modelling zeroday behaviour or for deriving new patterns and signatures that generalize beyond predefined attack types. Second, most of these datasets are traffic-oriented, they provide packets, flows, and labels, but do not expose the underlying IDPS rules or the contextual information that drives rule engineering. They lack structured CTI (e.g., MITRE ATT\&CK, CVE/CWE, D3FEND, MISP, OTX), RCRA, and detailed payload-level semantics such as reusable regular expressions or YARA-style content patterns. This also  prevents the systematic training and evaluation of LLMs on realistic rule-generation and rule-verification tasks. To address these limitations, we proposed GenTI to fill precisely this gap by providing a rule-centric, CTI-enriched, payload-aware benchmark that exposes both production rules and their surrounding context for LLM-based IDPS rule synthesis.
\section{Preliminary}
This section present IDPS  and YARA rule structure to understand rule architecture.  
IDPS uses signature-based rules that provide rudimentary payload inspection capabilities and mainly function at the network and transport levels (L3/L4). The four main parts of an IDPS rule are the rule header, rule message, rule content, and rule metadata, as shown in Fig. \ref{fig:0r}. The action type (drop/reject for active IPS prevention or alert for detection-only IDS mode), source and destination addresses with port numbers, protocol specification, and traffic direction are among the network-layer attributes defined by the rule header. Simple payload matching, such as matching HTTP methods or file extensions (e.g., \texttt{content="GET"; content:".exe"}) is made possible by the rule content section using string patterns and regular expressions. Nevertheless, the payload structure's spatial precision is lacking in these content matches. The throughput speed is given priority in this architectural design, IDPS engines like Snort and Suricata can process several million packets every second at network perimeters, making them appropriate for real-time traffic monitoring and quick threat alerts based on recognized protocol characteristics.

\begin{figure}[htbp]
    \centering
    \includegraphics[scale=0.20]{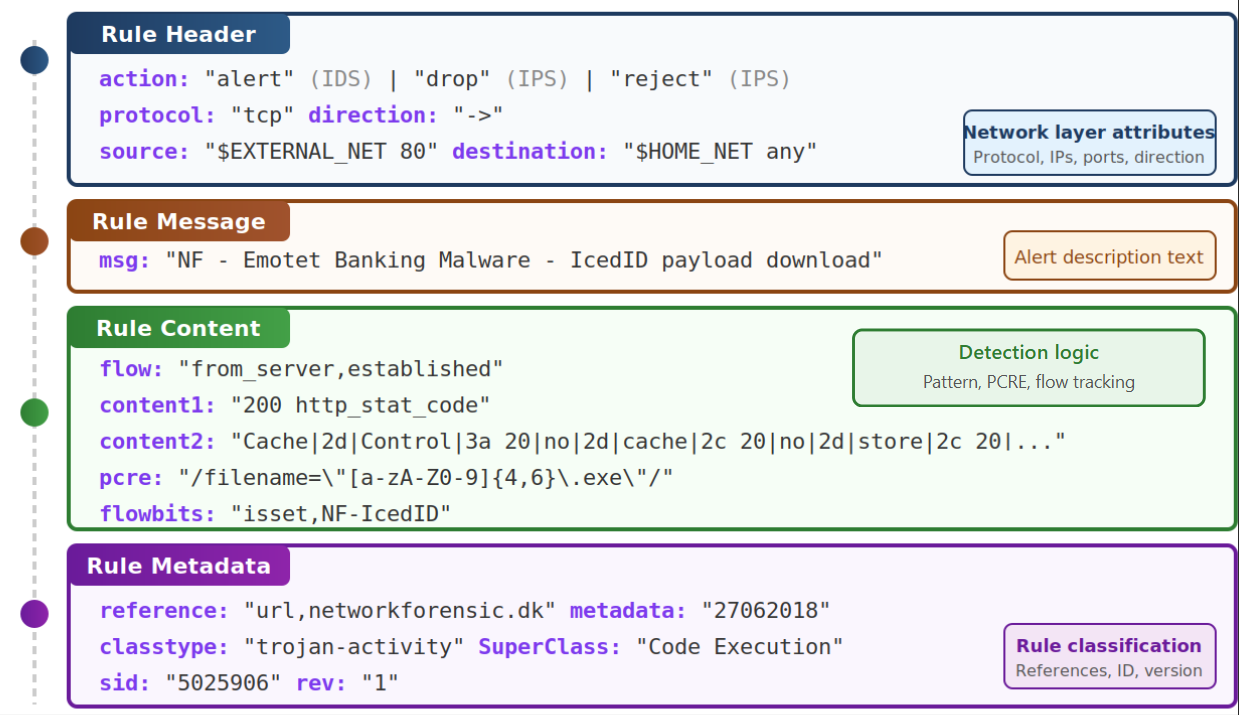}
    \caption{Modern IDPS Rule Structure}
    \label{fig:0r}
\end{figure}

Unlike header-level IDPS rules, YARA offers a pattern-matching framework made especially for deeper payload analysis at the application layer (L7), allowing accurate binary analysis of memory dumps, network streams, and file contents. A YARA rule, as shown in Fig. \ref{fig:02r}, is composed of four sections: a strings section that defines payload-level signatures, a condition section that specifies the detection logic, a meta section that documents threat intelligence context, containing malware family attribution, and the rule header that contains the rule name and classification tags. Hexadecimal byte sequences with wildcards (such as \texttt{\{6A 40 68 00 30 00 00\}}), ASCII and Unicode text strings, and regular expressions to match Command \& Control (C2) communication patterns are all supported by the strings section. Crucially, offset-precise matching is made possible by the condition section, which enables analysts to provide specific byte locations (like \verb|\$mz at 0| for PE header validation) or byte ranges inside file structures (like \verb|$shellcode in (0x400..0x1000)| for code section analysis). By identifying particular shellcode patterns, decryption techniques, and API call sequences at their anticipated locations inside executable binaries, this specific positioning capability enables precise malware family. 
\begin{figure}[htbp]
    \centering
    \includegraphics[scale=0.12]{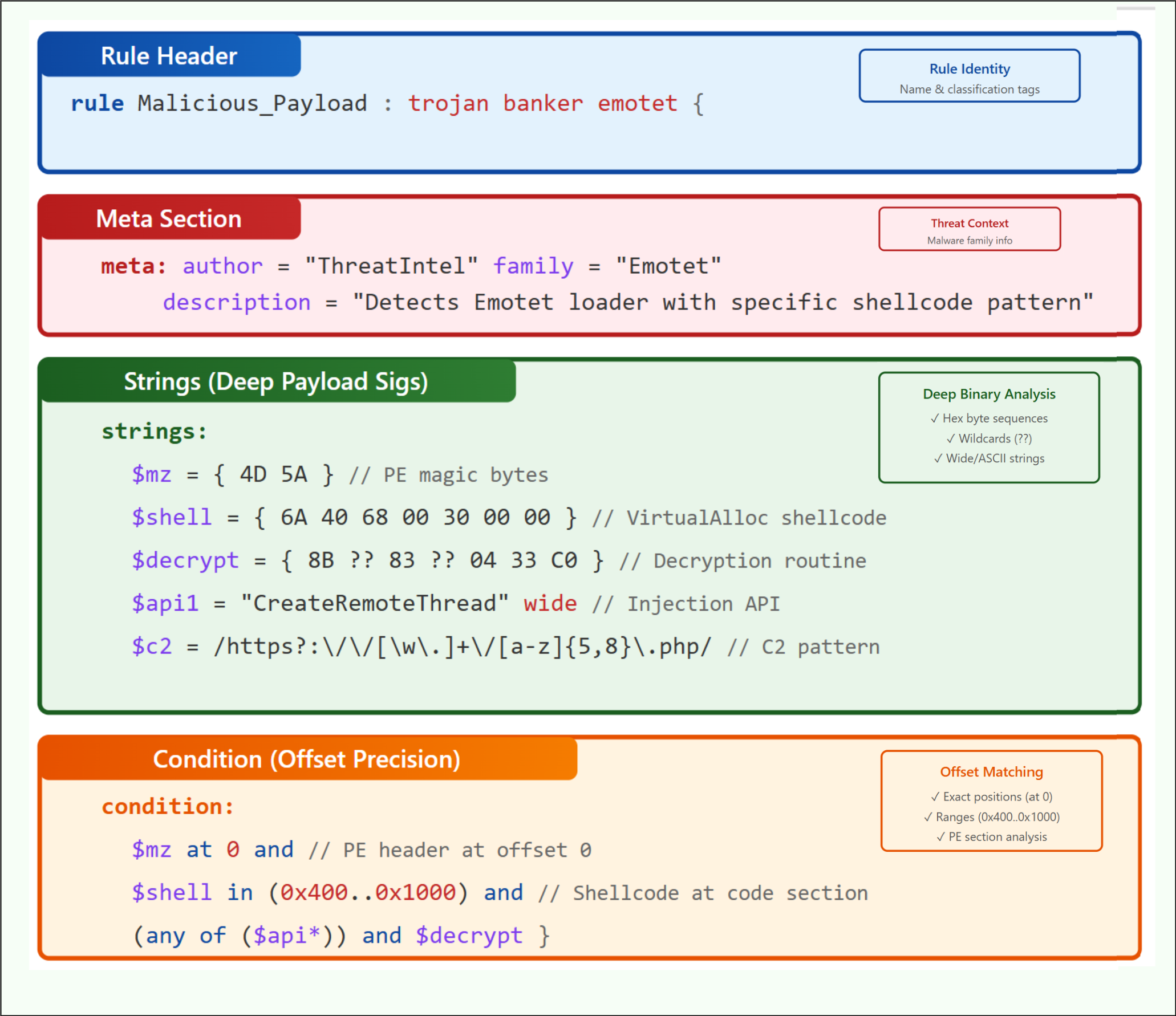}
    \caption{YARA Rule Structure}
    \label{fig:02r}
\end{figure}
\vspace{-1ex}
\section{Methodology for creating the dataset} \label{sec:data}
\begin{table*}[t]
\centering
\scriptsize
\caption{GTI Field Groups and Coverage}
\begin{tabular}{p{3.5cm} p{7.5cm} p{1.5cm} p{3cm}}
\hline
\textbf{Category} & \textbf{Fields} & \textbf{Coverage} & \textbf{Purpose} \\
\hline
Header Features (8) &
protocol, src\_ip, dst\_ip, src\_port, dst\_port, tcp\_flags, ttl, ip\_fragments &
100\% & Network packet analysis \\

Flow Features (3) &
flow\_direction, flow\_state, detection\_filter &
67\% & Connection tracking \\

Payload Features (3) &
payload\_content, application\_layer\_field, target\_service &
100\% & Deep packet inspection \\

Threat Context (10) &
classtype, attack\_type, mitre\_techniques, mitre\_tactics, mitre\_d3fend, mitre\_report, ioc, cve\_id, cwe\_id, malware\_family &
78\% & Threat intelligence \\

OTX Intelligence (4) &
otx\_pulses, otx\_cves, otx\_malware, otx\_refs &
75\% & AlienVault threat data \\

MISP Intelligence (8) &
misp\_event\_ids, misp\_threat\_level, misp\_tags, misp\_galaxies, misp\_sightings, misp\_last\_seen, misp\_org, misp\_attr\_ids &
88\% & Community threat sharing \\

Action \& Impact (3) &
action, severity\_level, confidence\_score &
100\% & Rule execution control \\

Training Prompts (6) &
rationale, reference, input\_prompt, user\_prompt, rule\_syntax, target\_output &
100\% & LLM training data \\

Rule Metadata (3) &
rule\_text, sid, rev &
100\% & Rule identification \\
\hline
\end{tabular}
\label{tab:autosecrules_idps_fields}
\end{table*}

\subsection{Datasets (GTI)}
We curate a CTI-enriched corpus, termed GTI, which is organised into two complementary components: an IDS/IPS rule component (GTI–IDPS) and a YARA rule component (GTI–YARA). The IDPS part captures network-level behaviour, while the YARA part targets payload and file-level patterns. Together, these components support the analysis as well as generation of rules that operate both on packet headers/flows and on content.

Further, the GTI–IDPS component is derived from Snort and Suricata rule sets together with selected emerging signatures. Each rule is normalised into a fixed schema that groups fields into header, flow, threat-context, CTI, and rule-metadata categories. At a high level, the dataset records how the rule views the traffic, including the protocol, direction, and application service. It also specifies what the rule is designed to detect  (such as payload patterns and attack types) and how it fits within the broader threat-intelligence landscape through MITRE ATT\&CK/D3FEND mappings, CVE/CWE links, and malware family associations. Additional blocks store intelligence from AlienVault OTX and MISP, as well as the operational action, severity with confidence level of the rule. Finally, a dedicated prompt block encodes an input prompt, user-style prompt, rationale, rule-syntax explanation and target rule, which we later use for LLM training and evaluation. The complete set of 48 structured fields and their categories is summarised in Table \ref{tab:autosecrules_idps_fields}.

Next, the GTI–YARA component focuses on payload-level signatures. Each YARA rule is represented by its identifier, rule body and parsed abstract syntax tree, along with the indicators embedded in the rule (metadata, string definition, condition expression, and registry keys). For every rule we record both counts and lists of these IoCs, together with CTI from MISP and OTX, mappings to ATT\&CK and D3FEND, quality scores and provenance information. Similar to the IDPS part, we attach LLM-oriented annotations such as an explanation of what the rule detects, recommended use-cases, deployment notes, suggested variants and safety warnings.

\subsection{Dataset Construction}
\label{subsec:dataset_construction}

The construction of the GTI dataset follows the CTI-enriched pipeline  as shown in Fig. \ref{fig:02}, which consists of twelve stages (steps 1--12) from rule ingestion to LLM-ready examples and response actions. The main steps are outlined below.
\begin{itemize}
    \item First, we aggregate base rules from multiple sources (steps 1--2). Snort and Suricata distributions and selected community rule sets provide the initial IDPS signatures. Besides, high-quality public YARA repositories and in-house rules contribute the payload-level signatures.
    After de-duplication as well as normalisation, we incorporated two additional repositories containing both \texttt{So\_Rules}  (shared-object rules) and \texttt{Preproc-Rules} (preprocessor-based detection rules)\footnote{\url{https://www.snort.org/downloads}}. These expanded rule sources were then passed to the GTI controller in the Data Source block. The controller orchestrates the flow of rules towards the parser, CTI modules, LLM models, and the database.

\begin{figure*}[htbp]
    \centering
    \includegraphics[scale=0.12]{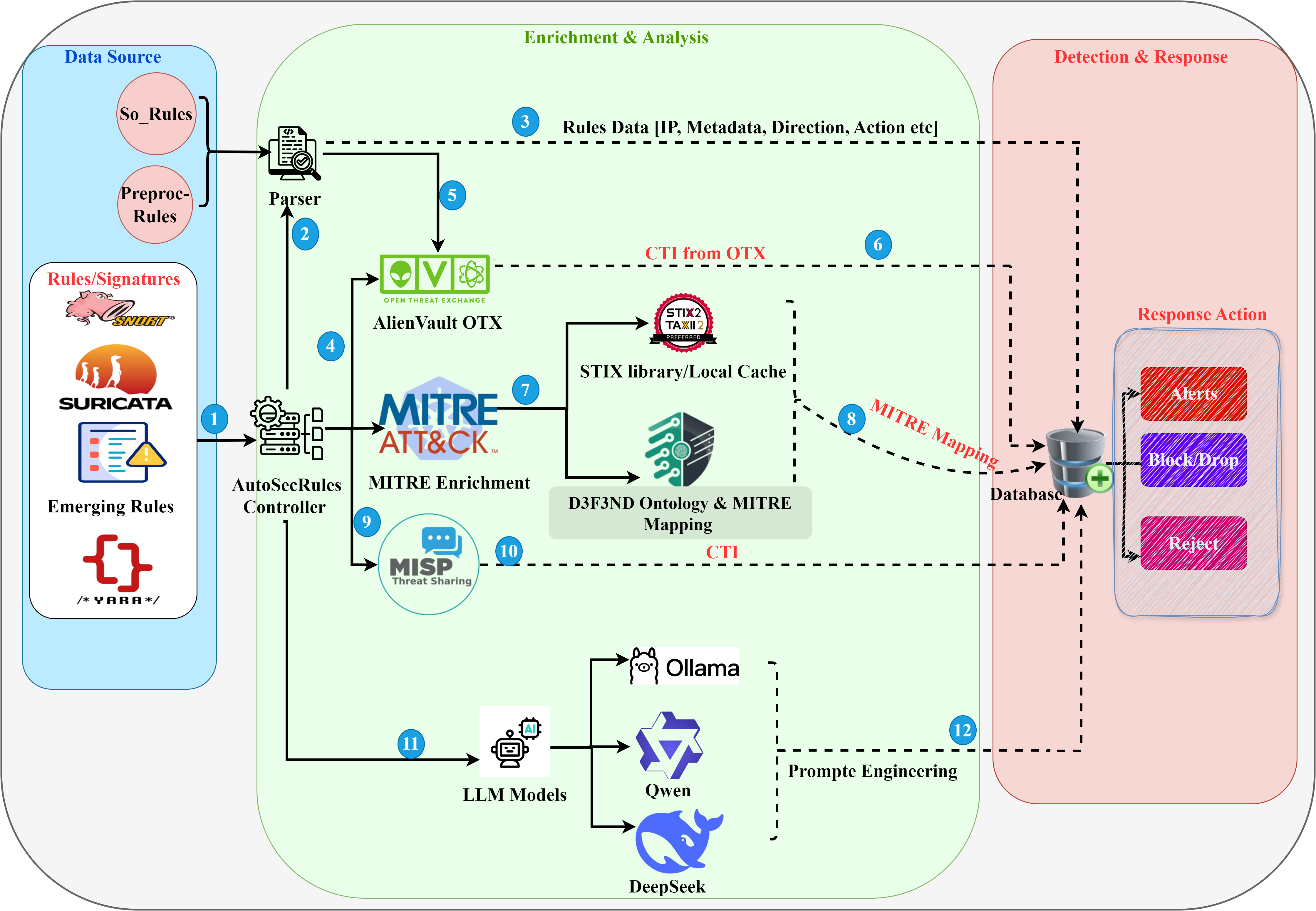}
    \caption{CTI-enriched dataset construction infrastructure for GenTI}
    \label{fig:02}
\end{figure*}

\item Second, we perform rule parsing and schema normalisation (step 3). For IDPS rules, a dedicated parser extracts header, flow features, such as the network protocol, source and destination IP addresses, along with ports, TCP flags, TTL, and fragmentation bits. It also identifies flow direction and state, detection filters, in addition to application-layer payload content (e.g., HTTP fields or target service). Together with this, we derive a compact initial context, including the \texttt{classtype} and high-level \texttt{attack\_type}. For YARA rules, we construct an abstract syntax tree (\texttt{rule\_ast}) from the raw rule body and isolate imports, metadata, string definitions and condition expressions. At this stage, each rule is represented as structured rules data (IP-level fields, metadata, direction and action) that can be systematically enriched and, where appropriate, directly stored in the repository.

 \item Third, we derive Indicators of Compromise (IoCs) from the parsed rules (step 4). In GTI--IDPS, each signature is associated with a primary \texttt{IoCs} and, where available, \texttt{cve\_id}, \texttt{cwe\_id} as well as \texttt{malware\_family}. In GTI--YARA, we count and enumerate all indicator types embedded in the rule body, including  metadata, string definition, condition expression, and registry keys (stored in the corresponding \texttt{iocs\_*} fields). These IoCs form the bridge between static rule content as well as external threat-intelligence services and are forwarded by the GTI controller to the MITRE enrichment modules.

 \item Fourth, we enrich both components with external CTI from AlienVault OTX (steps 5--6). The extracted IoCs are submitted to OTX to retrieve related pulses, CVEs, malware entries, and external references, which are stored in the \texttt{otx\_*} fields. The returned CTI is then normalised into STIX objects and cached in a local STIX/TAXII store. This intermediate STIX representation enables consistent reasoning over OTX data and avoids repeated external lookups when rules share indicators.

\item Fifth, we perform MITRE ATT\&CK and D3FEND mapping using the MITRE Enrichment module (steps 7--8). The STIX-encoded CTI, together with rule descriptions and any embedded references, is mapped to one or more ATT\&CK tactics as well as techniques,  in addition to relevant software and adversary groups. The same context is passed through the D3F3ND ontology to obtain corresponding defensive techniques and countermeasures. In the IDPS component these mappings appear in \texttt{mitre\_techniques}, \texttt{mitre\_tactics}, \texttt{mitre\_d3fend}, and  \\ \texttt{mitre\_report}. In the YARA component we additionally maintain explicit counts (e.g., \texttt{mitre\_\*\_count}, \texttt{d3fend\_\*\_count}) alongside the lists themselves. The resulting ATT\&CK/D3FEND mappings are persisted to the central database as part of the threat-context layer.

\item Sixth, we integrate community threat intelligence from MISP (steps 9--10). Using the same IoCs, the GTI controller queries one or more MISP instances and retrieves event identifiers, attributes, tags, galaxies, threat levels, sightings, last-seen timestamps, and the contributing organisation. These responses are normalised into the \texttt{misp\_*} fields and attached to each rule instance. Together with the OTX/STIX and MITRE/D3FEND blocks, this yields a multi-source CTI profile for every rule.

\item Seventh, we compute quality and operational metrics. For IDPS rules, we record the runtime action (alert, drop, or reject), the severity level (\texttt{severity\_level}), and a confidence score \\(\texttt{confidence\_score}). This score reflects how well the CTI supports the rule and how reliable the signature is expected to be in practice. We compute overall, completeness, complexity, coverage, and CTI-confidence scores for YARA rules. Provenance fields \texttt{provenance\_source\_file} and \texttt{provenance\_collection \\ \_timestamp} are populated to ensure traceability and reproducibility of the dataset, as well as to support later ablation studies.

\item Eighth, the system employs a multi-model LLM architecture to achieve robust and nearly accurate CTI enrichment of the security ruleset for those rules which have missing IOCs and or highly simple rules (step 11). So all the rules should have proper relative information. This process bifurcates based on the presence
of existing IOCs; where IOCs are present. Further, LLMs are
utilized for targeted contextual enrichment. Conversely,
rules lacking IOCs undergo synthetic CTI generation
primarily facilitated by the Qwen-7B model. To ensure
the generation of high-fidelity, contextually relevant CTI,
the LLaMA 3.2-7B model is strategically deployed for
sophisticated instructional prompting (prompt engineer-
ing). Furthermore, all intricate human language-based
tasks, including the generation of use cases, the building
of rationale, and linguistic enrichment for auto prompt
refinement, are delegated to the DeepSeek-7B model.
This model is guided by precise instructions derived from
the LLaMA 3.2-7B model.

 \item Ninth, we integrate a prompt-engineering layer to support LLM-based rule synthesis and explanation (step 12). A structured \texttt{input\_prompt} is constructed from the header, flow, payload, and CTI fields for every IDPS rule. A corresponding \texttt{user\_prompt} is written in the style of an analyst asking for a rule. The canonical Snort/Suricata signature is stored as \texttt{target\_output}, while \texttt{rationale} and \texttt{rule\_syntax} provide natural-language explanations of the detection logic. YARA rules use \texttt{llm\_rationale}, \texttt{llm\_use\_cases}, \texttt{llm\_deployment\_notes}, and \texttt{llm\_variants}  to capture the model’s reasoning, intended deployment scenarios, and suggested variants, while \texttt{llm\_safety\_warnings}  highlights any potential misuse. These artifacts constitute the prompt-engineering block stored alongside the CTI-enriched records.

\item Finally, all enriched records are stored in the GTI repository and linked to the downstream detection-and-response layer (alerts, block/drop, reject), forming the right-hand part of the pipeline. The resulting corpus combines network-level IDPS rules and payload-level YARA signatures with detailed CTI mappings, MITRE/D3FEND context, RCRA, quality metrics, and LLM-oriented annotations. This forms a comprehensive benchmark for automatic new rules generation and evaluation, as illustrated in Fig. \ref{fig:02}.
\end{itemize}
\section{Proposed GenTI Framework}
\label{sec:proposed_method}

In this section, we introduce the GenTI framework, which transforms CTI-aware analyst prompts and representative attack payloads into verified IDPS rules. GenTI  operates as an end-to-end pipeline that integrates prompt analysis, LLM-based rule synthesis, CTI-driven attack mapping, COT, iterative rule refinement, and a structured CoVe stage. The overall architecture of the framework is depicted in Fig. \ref{fig:03}.

\begin{figure*}[htbp]
    \centering
    \includegraphics[scale=0.13]{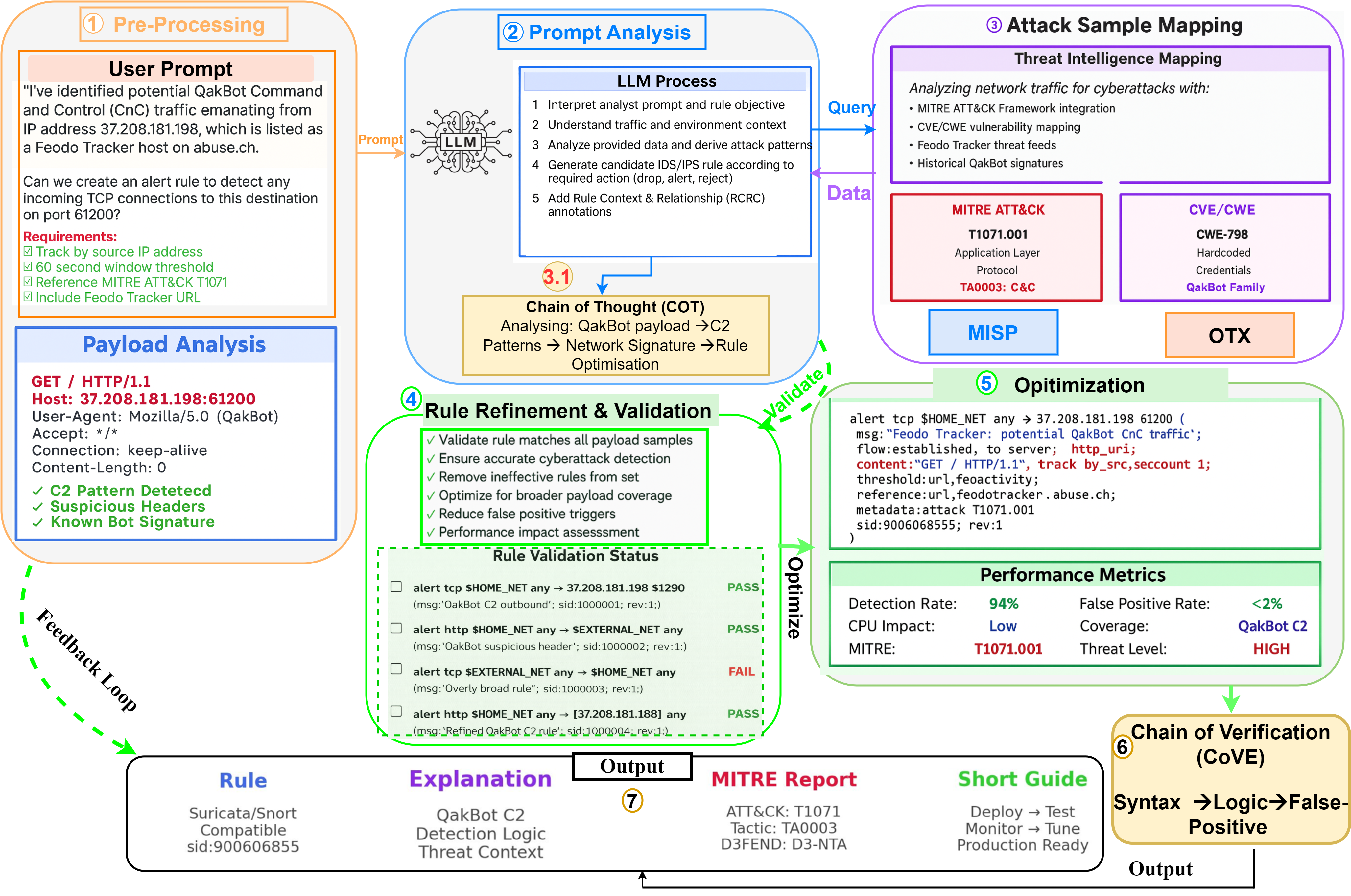}
    \caption{Proposed GenTI workflow for CTI-aware, LLM-driven intrusion detection rule generation and verification.}
    \label{fig:03}
\end{figure*}

\subsection{Solution Approach}

At a high level, GenTI takes as input a CTI-aware analyst prompt, representative malicious payload samples and the enriched rule context from the GTI dataset. The goal is to produce a IDPS compatible rule that is syntactically correct, semantically aligned with the analyst intent and mapped to the appropriate MITRE ATT\&CK and D3FEND entries, while also satisfying security-effectiveness constraints.

The \textbf{generation and optimisation phase} starts from structured context extracted from the analyst prompt and payloads. A prompt-construction module builds a structured prompt \(P_{\text{struct}}\) and an analyst-style user prompt \(P_{\text{user}}\). These are provided to the LLM together with selected CTI fields (e.g., mapped CTI techniques, IoCs, malware family) to generate an initial candidate rule \(r_0\) and a COT trace \(C_0\). 
GenTI then enters an iterative refinement loop, where each rule candidate \(r_k\) is evaluated on both attack and benign traffic. This evaluation produces a metric vector \(M_k\)
 that captures syntax validity, semantic similarity to the baseline intent, ATT\&CK coverage, detection rate, false-positive rate, and resource cost. If \(M_k\) does not satisfy configurable thresholds, a feedback message \(F_{k+1}\) is constructed and fed back into the LLM to produce a refined rule \(r_{k+1}\). This loop continues until convergence, as captured in the refinement block of Algorithm~\ref{alg:autosecrule}.

Next, the \textbf{verification phase} applies a structured CoVe pipeline to the best candidate from the optimisation phase. CoVe comprises three ordered checks. In the first phase, CoVe performs a syntactic check to ensure Snort/Suricata compatibility, including the header, options, and keyword structure. In the second phase, it conducts a logical check comparing the rule conditions with the mapped ATT\&CK technique, payload semantics, and CTI context. In the third phase, CoVe executes a false-positive analysis by replaying benign traffic profiles. If any check fails, the rule is rejected and may trigger another refinement cycle; only rules that pass all CoVe stages are promoted to final outputs. For each accepted rule, GenTI also computes a \emph{composite score} combining syntax accuracy, semantic similarity, CTI coverage and security effectiveness, which is later used in comparative evaluation.

\subsection{GenTI System Modules}

To implement the above solution, GenTI is decomposed into the following modules, corresponding to the numbered blocks in Fig.~\ref{fig:03}:

\begin{enumerate}
  \item \textbf{Analyst Prompt \& Payload Analysis (Pre-processing):}  
  The process begins with an analyst describing the suspected threat in natural language (e.g., QakBot C2 traffic to a known Feodo Tracker host). This description includes explicit requirements such as tracking by source IP, temporal thresholds, and references to specific ATT\&CK techniques. In parallel, representative payloads are parsed, and salient features such as HTTP headers, C2 markers and known bot signatures are extracted to form an initial feature set \(F_0\).

  \item \textbf{Prompt Construction and LLM Process (Prompt Analysis:}  
  The prompt-construction module consolidates the analyst intent, payload features and selected CTI attributes into a structured prompt \(P_{\text{struct}}\) and a natural-language prompt \(P_{\text{user}}\). The LLM process interprets the detection objective, infers the traffic and environment context, and analyses payload patterns. It then generates an initial IDPS rule candidate \(r_0\) along with RCRA and CoT  \(C_0\).
  \item \textbf{Threat-Intelligence Mapping (Attack Sample Mapping):}  
  Using indicators extracted from \(r_0\) and the payloads, this module queries the CTI store built from AlienVault OTX, MISP and MITRE ATT\&CK/D3FEND. It returns a threat-intelligence profile comprising technique and tactic identifiers (e.g., T1071.001 / TA0003), related CVE/CWE entries, malware families and external feed references. This profile is attached to the candidate rule as its threat context.

  \item \textbf{Rule Refinement and Validation Engine:}  
  This engine realises the refinement loop in Algorithm~\ref{alg:autosecrule}. Each candidate rule \(r_k\) is evaluated against malicious and benign traces and representative payload variants. Basic checks ensure that the rule fires on all known attack samples and remains silent on clean traffic. Rule variants that either over-generalise or miss important samples are discarded. The engine records status labels (PASS/FAIL) for each variant and generates focused feedback \(F_{k+1}\) that highlights missing patterns, overly broad conditions or conflicting options.

  \item \textbf{Optimisation and Metric Computation:}  
 Candidates that pass basic validation proceed to a performance evaluation stage, where their effectiveness and impact are assessed. In this stage, GenTI computes metrics such as detection rate, false-positive rate, computational impact (e.g., average CPU utilisation), coverage of the targeted ATT\&CK technique, and an overall threat level.

  \item \textbf{Chain-of-Verification (CoVe):}  
  CoVe performs final verification in three ordered stages: syntactic verification, logical verification and false-positive analysis. Syntactic verification ensures that \(r_k\) can be loaded by the IDS engine without errors. Logical verification checks the consistency between the rule and its ATT\&CK/D3FEND mapping. It validates that the protocol, ports, payload patterns, and flow direction align with the documented behaviour for the specific technique. False-positive analysis replays benign traffic profiles to identify noisy triggers; rules that cause unacceptable alert volumes are rejected.

  \item \textbf{Output Assembly and Reporting:}  
  Rules that pass CoVE are used by GenTI to construct a final output record. This record includes the verified Suricata/Snort rule with SID and revision, a natural-language explanation based on the CoT describing the QakBot C2 detection logic and its threat context, a concise MITRE report with associated techniques, tactics, and D3FEND defences, and a short deployment guide. The deployment guide summarises how the rule should be tested, monitored, and tuned before production use.
\end{enumerate}

\begin{algorithm}[t]
\scriptsize
\caption{GenTI: CTI-aware LLM rule generation and verification}
\label{alg:autosecrule}
\begin{algorithmic}[1]

\Require Analyst prompt $P$; payload samples $\mathcal{S}$; CTI store $\mathcal{C}$
\Ensure Verified IDPS rule $\hat{r}$ with context $\hat{c}$

\Statex
\State \textbf{Pre-processing and prompt construction}
\State $(P_{\text{struct}}, P_{\text{user}}, F_0)
       \gets \textsc{BuildPrompt}(P, \mathcal{S})$

\Statex
\State \textbf{Initial LLM generation}
\State $(r_0, C_0)
       \gets \textsc{LLMGenerate}(P_{\text{struct}}, P_{\text{user}}, F_0)$

\Statex
\State \textbf{Threat-intelligence mapping}
\State $c_0 \gets \textsc{MapCTI}(r_0, \mathcal{S}, \mathcal{C})$

\Statex
\State \textbf{Refinement and validation loop}
\State $k \gets 0$
\While{\textsc{NotConverged}$(r_k, c_k)$}
    \State $M_k \gets \textsc{EvaluateRule}(r_k, \mathcal{S})$
          \Comment{DR, FPR, CPU, coverage}
    \If{\textsc{MeetsThresholds}$(M_k)$}
        \State \textbf{break}
    \EndIf
    \State $F_{k+1} \gets \textsc{BuildFeedback}(M_k)$
    \State $(r_{k+1}, C_{k+1})
           \gets \textsc{LLMRefine}(P_{\text{struct}}, P_{\text{user}}, C_k, F_{k+1})$
    \State $c_{k+1} \gets \textsc{UpdateCTI}(r_{k+1}, \mathcal{C})$
    \State $k \gets k + 1$
\EndWhile

\Statex
\State \textbf{Chain-of-Verification (CoVe)}
\If{\textbf{not} \textsc{SyntacticOK}$(r_k)$}
    \State \Return \textsc{RejectRule}()
\EndIf
\If{\textbf{not} \textsc{LogicOK}$(r_k, c_k)$}
    \State \Return \textsc{RejectRule}()
\EndIf
\If{\textbf{not} \textsc{FPAnalysisOK}$(r_k)$}
    \State \Return \textsc{RejectRule}()
\EndIf

\Statex
\State \textbf{Finalisation}
\State $\hat{r} \gets r_k$
\State $\hat{c} \gets \textsc{BuildContext}(C_k, c_k)$
       \Comment{Explanation, MITRE report, short guide}
\State \Return $(\hat{r}, \hat{c})$

\end{algorithmic}
\end{algorithm}
\subsection{CoT/CoVE-based Data Augmentation}

The LLM component inside GenTI is trained not only on plain rule--prompt pairs, but also on auxiliary examples that encourage explicit reasoning and self-verification. We construct additional training samples for each rule instance in the GenTI dataset that expose the CoT and CoVe structure of the framework.

Concretely, given a CTI-enriched rule record, we generate a reasoning-style prompt that asks the model to explain why the rule correctly detects a given attack scenario. We also generate a verification-style prompt that asks the model to check the rule for syntactic errors, logical inconsistencies, and potential false-positive conditions. The corresponding targets are derived from the ground-truth rule, its CTI mappings (e.g., ATT\&CK technique, malware family) and pre-computed validation results. These augmented samples are stored as additional CoT and CoVe entries in the training set and are interleaved with standard generation examples during fine-tuning.  This augmentation has two effects. First it teaches the model to verbalise the detection logic and threat context that underlie a rule, which directly improves the quality of the natural-language explanations produced at inference time. Second, it exposes the model to structured verification patterns (syntax checks, semantic checks and FP analysis), thereby aligning the model's internal reasoning with the downstream CoVe pipeline and improving robustness to imperfect prompts.

\subsection{Curriculum Learning with Difficulty Levels}
\label{subsec:curriculum_training}

To further stabilise training and improve generalisation, we adopt a multi-stage curriculum learning strategy over the  dataset. Each rule instance is assigned a scalar \emph{difficulty score} that reflects the complexity of its structure and threat context. The score accounts for factors such as the number of header and payload constraints, the richness of CTI annotations (including multiple ATT\&CK techniques and CVE/CWE links), and the presence of CoT/CoVe augmented prompts. Lower scores correspond to simple header-only rules with minimal CTI context, whereas higher scores correspond to rules with complex payload patterns and dense CTI mappings.

We partition the training set into a sequence of curriculum stages, where early stages contain only low-difficulty examples and later stages progressively introduce more complex rules. In Stage~1, the model is exposed primarily to syntactically simple rules and basic generation prompts, which allows it to quickly learn the core Snort/Suricata syntax and common header patterns. Stage~2 introduces examples with richer CTI context and simple CoT prompts, encouraging the model to connect rule structure with ATT\&CK techniques and IoCs. Stage~3 adds harder CoVe-augmented examples, emphasising verification and robustness. Finally, Stage~4 combines the full difficulty range of rules. Each stage is trained for a fixed number of epochs with a stage-specific learning rate schedule, and the model parameters are warm-started from the previous stage.

This curriculum design prevents the model from overfitting to complex CTI-heavy rules at the beginning of training, while still allowing it to master the full spectrum of rule complexity by the final stage. In our experiments, the curriculum yields higher syntax accuracy, better semantic similarity and improved CTI coverage compared to a single-stage fine-tuning regime on the same data.

\section{Experiments and Evaluation} \label{sec:exp}

\subsection{Dataset (GTI) and Experimental Configuration}
\label{subsec:dataset_description}
In this section, we present the experimental evaluation of GenTI, relying on the CTI-enriched GTI corpus introduced in Section~\ref{sec:data}. The corpus
contains approximately $150\,\text{k}$ IDPS rules and
$50\,\text{k}$ payload-level YARA rules. The network subset aggregates signatures from Snort and Suricata taken from their official distributions and curated community feeds. It also includes a small number of high-quality emerging rules contributed by industry partners. Each IDPS rule is normalised into the GTI--IDPS schema
(Table~\ref{tab:autosecrules_idps_fields}). Every rule instance is annotated with an impact level (low, medium, high, critical) based on its mapped MITRE ATT\&CK technique, associated CVEs, and analyst-provided severity labels. This produces a distribution that is skewed toward high and critical threats (Fig. \ref{fig:04}). In addition, we assign each example a difficulty score that captures structural and contextual complexity, such as the number of payload constraints, richness of CTI annotations, and presence of CoT/CoVe-augmented prompts. This score is then used to build the curriculum schedule.
\begin{figure}[htbp]
    \centering
    \includegraphics[scale=0.16]{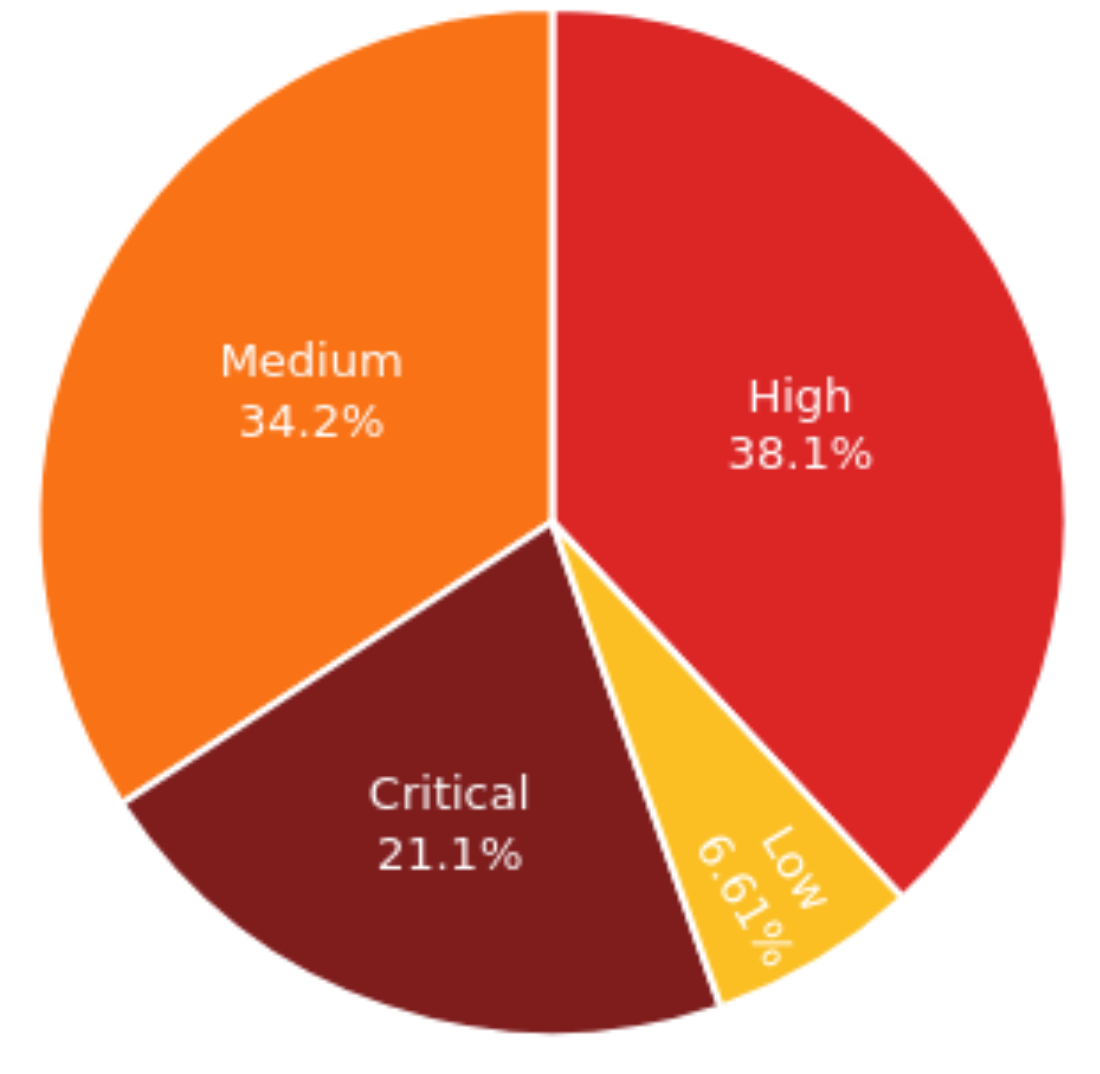}
    \caption{Attacks Severity Level}
    \label{fig:04}
\end{figure}

Next, the combined corpus is split in a stratified manner into 80\% training, 10\% validation, and 10\% test data. This preserves the distribution of rule sources, MITRE ATT\&CK techniques, severity levels, and difficulty scores, while ensuring that rules remain within the same partition. By default, all quantitative results in this section are reported on the held-out test set. GenTI is implemented in PyTorch using HuggingFace Transformers
and fine-tuned on a  8 Nividia A100 GPUs. We adopt a QLoRA-based recipe on the DeepSeek-7B backbone, training lightweight adapters over four curriculum stages that progress from simple header/payload patterns, through complex rule-syntax learning and reasoning (e.g., evasion and pattern detection), to CTI-rich, verification-oriented examples. The mini-batch size, learning rate of $10^{-4}$, and the number of epochs per stage are chosen to fit within the A100 memory budget while maintaining high GPU utilisation. Early stopping is applied based on the validation composite score.

\subsection{Evaluation Metrics}
\label{subsec:evaluation_metrics}

We evaluate GenTI on a held-out test set of CTI-aware prompts and
their corresponding expert-analyst rules. For each test instance the
model generates a candidate rule, which is then scored along five
dimensions, Syntax Accuracy (SA), Semantic Similarity (SS), CTI Coverage (CC), Security Effectiveness (SE), and a Composite Score (CS). All metrics are
normalised to the range $[0,1]$ unless stated otherwise.

\subsubsection{Syntax Accuracy (SA)}
Syntax accuracy measures whether the generated rules are accepted by
the target engine without manual correction. Let $N$ be the number of
generated rules and ${\bf 1}\{\cdot\}$ the indicator function that
returns~1 when its argument is true and~0 otherwise. A rule
$\hat{r}_i$ is considered syntactically valid if it can be loaded by
Snort/Suricata (or the YARA engine) without parse errors:
\begin{equation}
\mathrm{SA} = \frac{1}{N} \sum_{i=1}^{N}
{\bf 1}\{\textsc{ParseOK}(\hat{r}_i)\}.
\end{equation}
This metric captures the model's ability to respect low-level rule
syntax and keyword structure.

\subsubsection{Semantic Similarity (SS)}
Semantic similarity evaluates how closely a generated rule
$\hat{r}_i$ matches the corresponding reference rule $r_i$ in terms of
detection intent. We decompose each rule into a set of key fields
(e.g., protocol, direction, source/destination ports, flow options,
main content pattern, action). Let $\mathcal{F}$ denote this field set
and $f(r)$ the value of field $f$ in rule $r$. The per-instance
similarity is defined as the fraction of fields that match:
\begin{equation}
\mathrm{SS}_i = \frac{1}{|\mathcal{F}|}
\sum_{f \in \mathcal{F}} {\bf 1}\{ f(\hat{r}_i) = f(r_i) \},
\end{equation}
and the overall semantic similarity is the average over the test set:
\begin{equation}
\mathrm{SS} = \frac{1}{N} \sum_{i=1}^{N} \mathrm{SS}_i.
\end{equation}
For YARA rules we apply the same definition using protocol-agnostic
fields such as condition structure and sets of IoCs.

\subsubsection{CTI Coverage (CC)}
CTI coverage quantifies how well the generated rules preserve the
ground-truth ATT\&CK mappings. Let $T_i$ be the set of ATT\&CK
techniques associated with the reference rule $r_i$ and
$\hat{T}_i$ the set predicted (or implied) by the generated rule
$\hat{r}_i$. We define a per-instance Jaccard similarity:
\begin{equation}
\mathrm{CC}_i = 
\frac{|T_i \cap \hat{T}_i|}{|T_i \cup \hat{T}_i|},
\end{equation}
and report the average CTI coverage as
\begin{equation}
\mathrm{CC} = \frac{1}{N} \sum_{i=1}^{N} \mathrm{CC}_i.
\end{equation}
\subsubsection{Security Effectiveness (SE)}
Security effectiveness reflects the operational quality of the
generated rules when deployed on network traffic. For a given rule set,
we replay labelled malicious and benign traces and compute the
detection rate (true positive rate) and false positive rate:
\begin{equation}
\mathrm{DR} = \frac{\mathrm{TP}}{\mathrm{TP} + \mathrm{FN}}, \quad
\mathrm{FPR} = \frac{\mathrm{FP}}{\mathrm{FP} + \mathrm{TN}}.
\end{equation}
To obtain a single scalar score we combine these quantities as
\begin{equation}
\mathrm{SE} = \alpha \, \mathrm{DR} + (1-\alpha) (1 - \mathrm{FPR}),
\end{equation}
where $\alpha \in [0,1]$ controls the trade-off between detection and
false positives (we set $\alpha = 0.5$ in our experiments). When
available, we additionally monitor the normalised CPU utilisation
$\mathrm{CPU}$ of the NIDS engine and report it separately.

\subsubsection{Composite Score (CS)}
Finally, we report a composite score that aggregates the previous
dimensions into a single metric:
\begin{equation}
\mathrm{CS} = 
w_{\mathrm{SA}} \, \mathrm{SA} +
w_{\mathrm{SS}} \, \mathrm{SS} +
w_{\mathrm{MC}} \, \mathrm{MC} +
w_{\mathrm{SE}} \, \mathrm{SE},
\end{equation}
with $w_{\mathrm{SA}} + w_{\mathrm{SS}} + w_{\mathrm{MC}} +
w_{\mathrm{SE}} = 1$. In our default configuration we use
$w_{\mathrm{SA}} = 0.2$, $w_{\mathrm{SS}} = 0.3$,
$w_{\mathrm{MC}} = 0.2$ and $w_{\mathrm{SE}} = 0.3$, giving slightly
higher weight to semantic fidelity and security effectiveness. The
composite score is used to compare different model variants and
training strategies.
\begin{figure*}[htbp]
    \centering
    \includegraphics[scale=0.26]{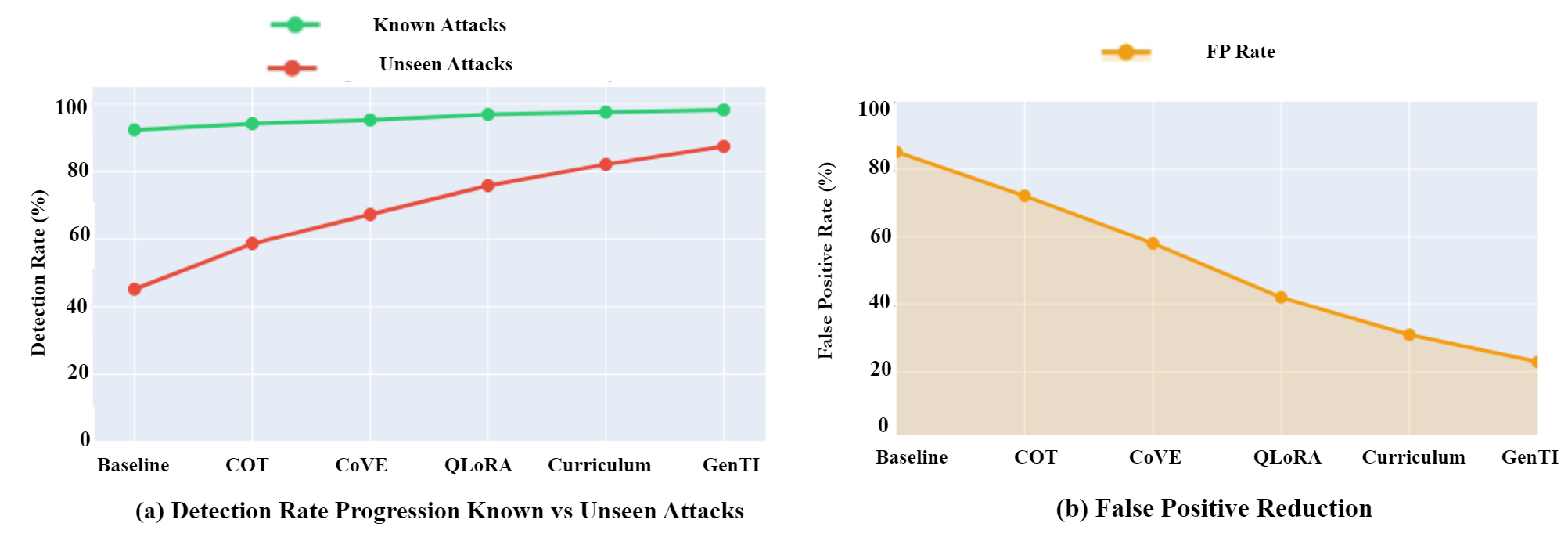}
    \caption{Unseen Attacks  Detection Performance Analysis}
    \label{fig:05}
\end{figure*}
\subsection{Cross-Model Evaluation on GenTI }

To evaluate the effectiveness of GenTI relative to baseline
LLMs, Table~\ref{tab:model_performance_complete} reports cross-model
performance on the GenTI test set in terms of SA, SS, CC, SE and CS. The zero-shot
GPT\mbox{-}3.5\mbox{-}Turbo baseline serves as a strong yet generic
reference point, obtaining SA~$\approx=0.72$, SS~$\approx=0.68$ and CC~$\approx=0.75$, with
a CS~$\approx=0.71$. Applying few-shot prompting to LLaMA‑3‑8B and Mistral‑7B produces consistent gains across all metrics, reaching composite score CC~$\approx=69.5\%$, CS~$\approx=73.0$. This indicates that general-purpose LLMs can benefit significantly from limited task-specific context when generating intrusion detection rules.

\begin{table}[htbp]
\scriptsize
\centering
\caption{Overall performance of GenTI  compared with baseline LLMs}
\label{tab:model_performance_complete}
\renewcommand{\arraystretch}{1.5}

\begin{tabular}{p{2.0cm} p{0.6cm} p{0.6cm} p{0.6cm} p{0.6cm} p{0.6cm} p{0.6cm}}
\toprule
\textbf{Model} & \textbf{SA} & \textbf{SS} & \textbf{CC} & \textbf{SE} & \textbf{CS} & \textbf{Params} \\
\midrule
Mistral-7B & 70.5\% & 67.9\% & 73.3\% & 66.3\% & 69.5\% & 7B \\
GPT-3.5 Turbo & 72.1\% & 68.6\% & 75.4\% & 70.3\% & 71.5\% & 175B \\
Phi-3-Medium & 71.4\% & 68.7\% & 74.1\% & 67.2\% & 70.1\% & 14B \\
LLaMA-3-8B & 72.8\% & 69.9\% & 75.9\% & 68.5\% & 71.6\% & 8B \\
Mixtral-8x7B & 73.5\% & 70.6\% & 76.8\% & 69.2\% & 72.3\% & 45B \\
LLaMA-3.1-8B & 74.1\% & 71.4\% & 77.1\% & 70.5\% & 73.0\% & 8B \\
DeepSeek-7B & 74.6\% & 72.5\% & 77.3\% & 70.7\% & 73.5\% & 7B \\
Qwen2-7B & 75.4\% & 72.6\% & 78.5\% & 71.3\% & 74.4\% & 7B \\
Gemini-1.0-Pro & 75.9\% & 73.4\% & 79.0\% & 72.1\% & 74.9\% & Unknown \\
Claude-3-Opus & 77.1\% & 74.5\% & 80.2\% & 73.2\% & 76.1\% & 175B \\
GPT-4 & 78.3\% & 75.8\% & 80.9\% & 74.4\% & 77.3\% & 1.76T \\
Qwen2.5-7B & 78.7\% & 76.2\% & 81.4\% & 74.8\% & 78.0\% & 7B \\
Gemini-1.5-Pro & 79.1\% & 76.9\% & 82.1\% & 75.4\% & 78.4\% & Unknown \\
GPT-4-Turbo & 79.9\% & 77.3\% & 82.9\% & 75.9\% & 79.0\% & Unknown \\
Claude-3.5-Sonnet & 80.5\% & 78.1\% & 83.7\% & 76.8\% & 79.8\% & 175B \\
GPT-4o & 81.2\% & 78.9\% & 84.1\% & 77.3\% & 80.4\% & Unknown \\
\midrule
\textbf{GenTI} & \textbf{89.5\%} & \textbf{87.2\%} & \textbf{94.8\%} & \textbf{85.9\%} & \textbf{89.4\%} & \textbf{7B} \\
\bottomrule
\end{tabular}
\end{table}

The proposed GenTI model,  trained with QLoRA, curriculum learning, and CoT/CoV-based augmentation, outperforms all baselines on every metric. It achieves this despite having the same parameter budget as other open-source models and far fewer parameters than GPT-4-Turbo.
 GenTI attains SA~$=89.5\%$, SS~$=87.2\%$, CC~$=94.8\%$ and SE~$89.4$, resulting in the highest
composite score of $89.4\%$. These results show that CTI-enriched supervision and the proposed training strategy improve not only syntactic correctness but also semantic alignment with expert labels. They also lead to rules that better cover the underlying MITRE ATT\&CK techniques.

\subsection{Unseen Attacks Detection Analysis}
\label{subsec:zero_day}

To assess how GenTI  generalises beyond seen threats, we measure
its ability to detect previously unseen attacks on a held-out
PCAP set \footnote{\url{https://www.malware-traffic-analysis.net/}}. Fig. \ref{fig:05} summarises the evolution of
detection performance across the training stages (Baseline, +CoT, +CoV,
+QLoRA, +Curriculum +GenTI). In Fig. \ref{fig:05} (a), the detection rate for known attacks increases moderately from about 92\% at baseline to 98.2\% for GenTI. This shows that the additional training components do not compromise performance on in-distribution traffic. In contrast, unseen attacks detection
improves much more sharply, from roughly $45\%$ at baseline to $87.4\%$
in the GenTI, corresponding to a gain of about $42$ percentage points. This gap illustrates that CoT/CoV augmentations and curriculum
training are particularly effective in teaching the model to recognise
attack patterns that were not explicitly present in the training rules.
\begin{figure*}[htbp]
    \centering
    \includegraphics[scale=0.20]{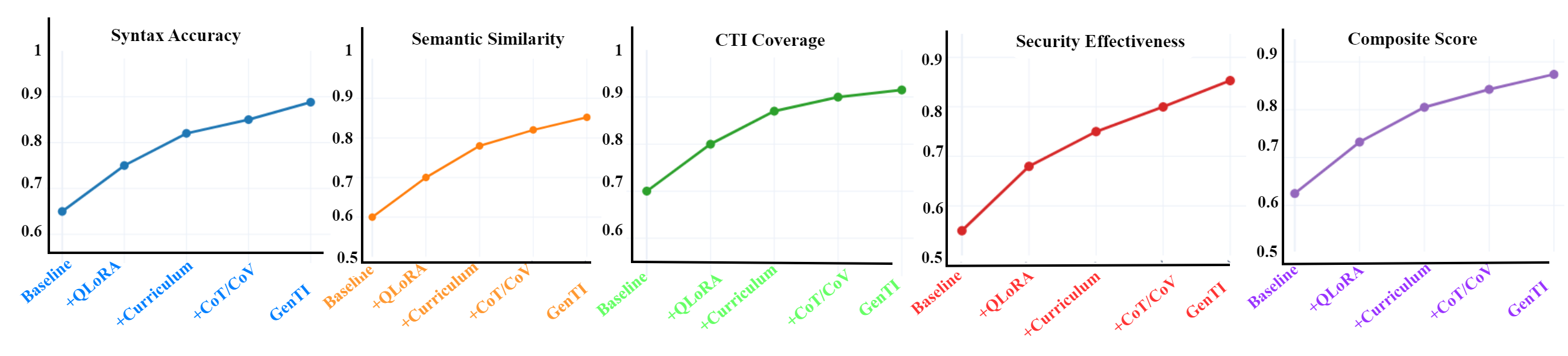}
    \caption{Ablation study of detection rule generation steps across evaluation metrics }
    \label{fig:08}
\end{figure*}

\begin{figure}[htbp]
    \centering
    \includegraphics[scale=0.29]{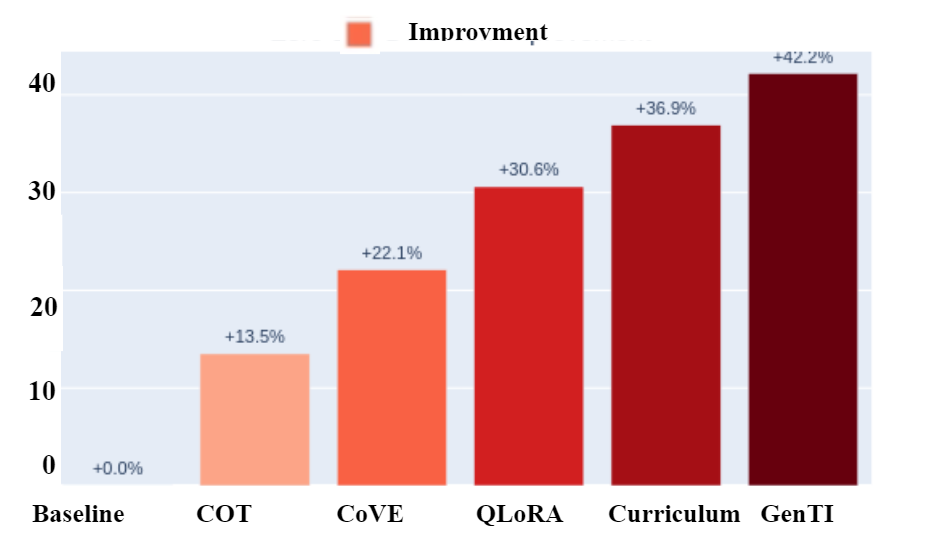}
    \caption{Improvement in Unseen Attacks Threat Detection}
    \label{fig:06}
\end{figure}
Fig.~\ref{fig:05} (b) reports the False-Positive (FP)
rate over the same stages. While unseen attacks detection increases
substantially, the FP rate decreases from around $8.5\%$ at baseline to
approximately $2.3\%$ in the final model. This monotonic reduction
suggests that the additional reasoning and verification signals help the
model refine decision boundaries rather than simply making its rules
more aggressive. The unseen attacks improvement is further shown in Fig.~\ref{fig:06}, where the relative gain increases from 0\% at baseline to more than 42\% for the final configuration. This highlights the combined contribution of CoT, CoV, QLoRA, and the curriculum schedule.

\begin{figure}[htbp]
    \centering
    \includegraphics[scale=0.31]{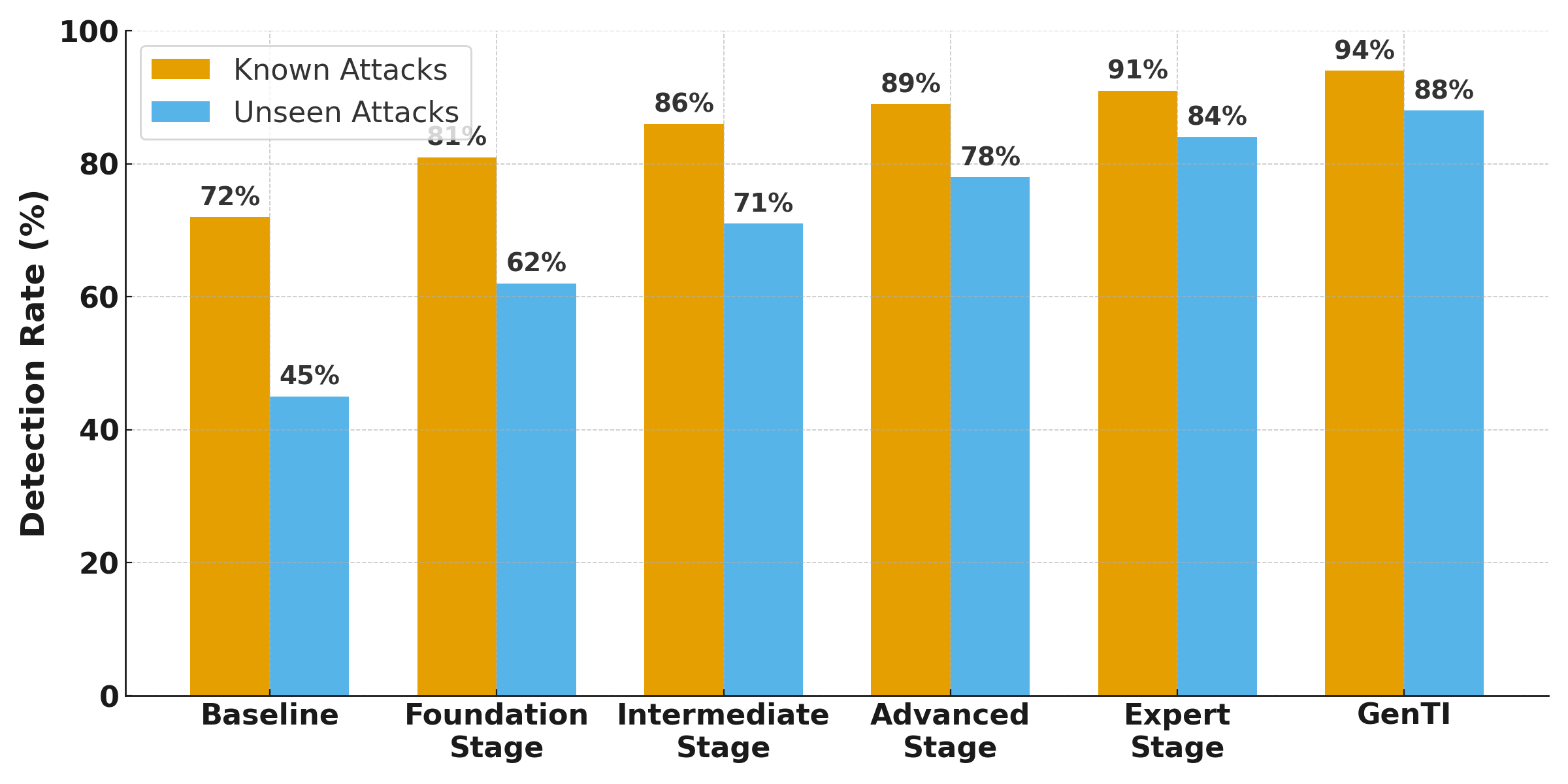}
    \caption{Detection Rate Comparison for known and Unseen Attacks}
    \label{fig:07}
\end{figure}
Finally, Fig.~\ref{fig:07} compares the final
detection rates for known and unseen attacks side by side. Although
unseen attacks performance remains slightly below that for known attacks
($88.4\%$ versus $94.2\%$), the gap is substantially narrower than at
baseline and is achieved with a significantly lower FP rate. Overall, these results show that GenTI CTI-aware prompting, CoT/CoV feedback loop, and difficulty-aware curriculum collectively improve the model’s robustness to Unknown attacks. At the same time, they also enhance the model’s precision on benign traffic.

\subsection{Ablation Study of Training Components}
\label{subsec:ablation}

To evaluate the contribution of each training component in the
GenTI  pipeline, we conduct an ablation study whose results are
summarised in Fig~\ref{fig:08}. These graphs present five evaluation metrics (SA, SS, CC, SE, and CS) across five configurations: the baseline,+QLoRA, +Curriculum, Base+QLoRA+Curriculum+CoT/CoV, and the final GenTI system.

\begin{figure}[htbp]
    \centering
    \includegraphics[scale=0.28]{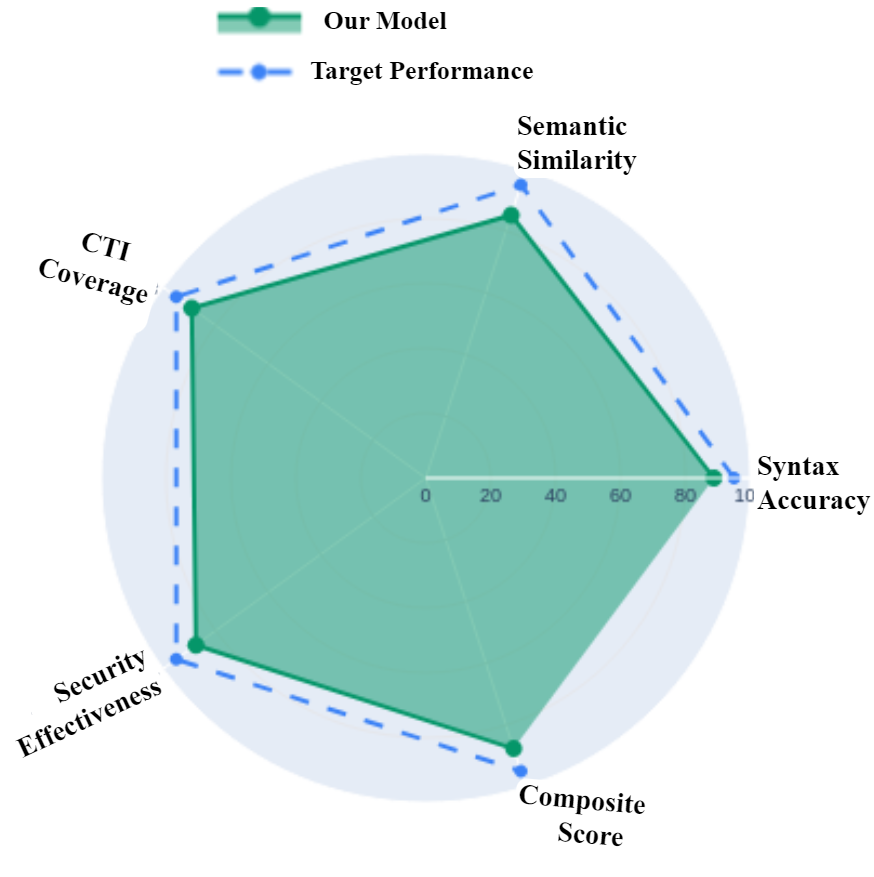}
    \caption{Final Performance Analysis}
    \label{fig:09}
\end{figure}

The Base LLM configuration, which relies solely on prompting without
any task-specific tuning, achieves moderate scores (e.g.,
SA~$\approx 0.65$, SS~$\approx 0.60$, CC~$\approx 0.70$). Introducing parameter-efficient fine-tuning with QLoRA yields the largest single improvement, with all five metrics increasing by roughly $8$–$13$ percentage points. This shows that alignment with the CTI-enriched GenTI corpus is essential for high-quality rule generation. Adding
difficulty-aware curriculum learning further boosts performance,
particularly in SS and SE, as the
model is gradually exposed to more complex, CTI-rich rules. Finally,
injecting CoT and CoVe signals produces a
consistent, though more incremental, gain across all metrics, mainly by
refining rule coverage and reducing residual errors. The resulting full configuration, GenTI  attains the highest scores on every metric. This demonstrates that QLoRA provides the main benefit, while curriculum learning and CoT/CoV work together to stabilise and refine the model’s behaviour.

Fig.~\ref{fig:09} provides an aggregated view of
GenTI final performance across all evaluation dimensions. The
radar plot contrasts the achieved scores (solid polygon) with a target
envelope (dashed polygon) that represents our desired operating range
for production deployment. GenTI closely approaches this target
on all axes, reaching high values for SA and SS (around 0.88), while maintaining strong CC above 0.9 and  SE in the mid-0.8 range. The
CS, which jointly accounts for these dimensions, also
remains close to the target, indicating that gains in one metric are
not obtained at the expense of others.
\vspace{-1ex}
\section{Conclusion}\label{sec:cn}

In this paper, we presented GenTI, a CTI-aware LLM framework that turns analyst prompts and attack payloads into verified IDPS and YARA rules. 
Most existing work reproduces traditional Snort/Suricata signatures and reports improvements in rule syntax or generic similarity. In contrast, GenTI is designed to extend rulebases with new, threat-intelligence–grounded rules, especially for unseen attacks. To support this, we curated the GenTI corpus (150k IDPS and 50k YARA rules), defined multi-view evaluation metrics ( SA, SS, CC, SE, CS), and coupled LLM-based rule generation with CTI mapping, iterative refinement and a CoVe loop.
Additionally, our experiments indicated that GENTI consistently outperforms strong baselines such as GPT-4o, LLaMA-3-7B, and Mistral-7B across all evaluation dimensions. It also achieved substantial improvements in unseen attacks detection while reducing false positives to operationally acceptable levels.

Our experiments show that GenTI consistently outperforms strong baselines such as GPT-4o, LLaMA-3-7B, DeepSeek-7B and Mistral-7B on all evaluation dimensions, while achieving substantial gains in unseen attacks detection and reducing false positives to operationally acceptable levels. The ablation study confirmed that QLoRA fine-tuning, curriculum training and CoT/CoV-based augmentation are all necessary to reach this robustness. Overall, GENTI demonstrated that LLMs can move beyond simply copying existing signatures and instead act as CTI-aware agents for expanding and maintaining IDS rulebases against evolving real-world threats.

\section*{Data Availability}
The dataset and code used in this study are available at \url{https://figshare.com/s/f34cd4706de24eecf0d6}
\bibliographystyle{IEEEtran}
\bibliography{cas-refs}

@inproceedings{wei2023xnids,
  title={$\{$xNIDS$\}$: Explaining deep learning-based network intrusion detection systems for active intrusion responses},
  author={Wei, Feng and Li, Hongda and Zhao, Ziming and Hu, Hongxin},
  booktitle={32nd USENIX Security Symposium (USENIX Security 23)},
  pages={4337--4354},
  year={2023}
}

@article{winkler2025proactive,
  title={Proactive threat detection in enterprise systems using Wazuh: A MITRE ATT\&CK Evaluation},
  author={Winkler, Aidan M and Sharma, Prinkle},
  journal={Computers \& Security},
  pages={104702},
  year={2025},
  publisher={Elsevier}
}

@article{ofte2023understanding,
  title={Understanding situation awareness in SOCs, a systematic literature review},
  author={Ofte, H{\aa}vard Jakobsen and Katsikas, Sokratis},
  journal={Computers \& Security},
  volume={126},
  pages={103069},
  year={2023},
  publisher={Elsevier}
}

@article{hadi2025uav,
  title={UAV-NIDD: A Dynamic Dataset for Cybersecurity and Intrusion Detection in UAV Networks},
  author={Hadi, Hassan Jalil and Cao, Yue and Khan, Muhammad Khurram and Ahmad, Naveed and Hu, Yulin and Fu, Chao},
  journal={IEEE Transactions on Network Science and Engineering},
  year={2025},
  publisher={IEEE}
}

@inproceedings{zhang2020cmirgen,
  title={CMIRGen: Automatic signature generation algorithm for malicious network traffic},
  author={Zhang, Runzi and Tong, Mingkai and Chen, Lei and Xue, Jianxin and Liu, Wenmao and Xie, Feng},
  booktitle={2020 IEEE 19th International Conference on Trust, Security and Privacy in Computing and Communications (TrustCom)},
  pages={736--743},
  year={2020},
  organization={IEEE}
}

@article{he2023adversarial,
  title={Adversarial machine learning for network intrusion detection systems: A comprehensive survey},
  author={He, Ke and Kim, Dan Dongseong and Asghar, Muhammad Rizwan},
  journal={IEEE Communications Surveys \& Tutorials},
  volume={25},
  number={1},
  pages={538--566},
  year={2023},
  publisher={IEEE}
}

@article{al2024analysis,
  title={Analysis of extreme learning machines (ELMs) for intelligent intrusion detection systems: a survey},
  author={Al-Haija, Qasem Abu and Altamimi, Shahad and AlWadi, Mazen},
  journal={Expert Systems with Applications},
  volume={253},
  pages={124317},
  year={2024},
  publisher={Elsevier}
}

@article{aldweesh2020deep,
  title={Deep learning approaches for anomaly-based intrusion detection systems: A survey, taxonomy, and open issues},
  author={Aldweesh, Arwa and Derhab, Abdelouahid and Emam, Ahmed Z},
  journal={Knowledge-Based Systems},
  volume={189},
  pages={105124},
  year={2020},
  publisher={Elsevier}
}

@misc{kddcup1999,
  author       = {{UCI KDD Archive}},
  title        = {KDD Cup 1999 Data},
  year         = {1999},
  howpublished = {\url{https://kdd.ics.uci.edu/databases/kddcup99/kddcup99.html}},
  note         = {Accessed: 2026-01-13}
}

@misc{nslkdd,
  author       = {{Canadian Institute for Cybersecurity}},
  title        = {NSL-KDD Dataset},
  year         = {2009},
  howpublished = {\url{https://www.unb.ca/cic/datasets/nsl.html}},
  note         = {Accessed: 2026-01-13},
  urldate      = {2026-01-13}
}

@misc{defconctfarchive,
  author       = {{DEF CON Communications, Inc.}},
  title        = {DEF CON Hacking Conference: Capture the Flag Archive},
  year         = {2024},
  howpublished = {\url{https://defcon.org/html/links/dc-ctf.html}},
  note         = {Accessed: 2026-01-13},
  urldate      = {2026-01-13}
}

@misc{caida2007ddos,
  author       = {{CAIDA}},
  title        = {The CAIDA {DDoS} Attack 2007 Dataset},
  year         = {2020},
  month        = jun,
  howpublished = {\url{https://www.caida.org/catalog/datasets/ddos-20070804_dataset/}},
  note         = {Accessed: 2026-01-13},
  urldate      = {2026-01-13}
}

@article{al2025multi,
  title={Multi-Stage Enhanced Zero Trust Intrusion Detection System for Unknown Attack Detection in Internet of Things and Traditional Networks},
  author={Al-Zewairi, Malek and Almajali, Sufyan and Ayyash, Moussa and Rahouti, Mohamed and Martinez, Fernando and Quadar, Nordine},
  journal={ACM Transactions on Privacy and Security},
  year={2025},
  publisher={ACM New York, NY}
}

@article{sharafaldin2018toward,
  title={Toward generating a new intrusion detection dataset and intrusion traffic characterization.},
  author={Sharafaldin, Iman and Lashkari, Arash Habibi and Ghorbani, Ali A},
  journal={ICISSp},
  volume={1},
  pages={108--116},
  year={2018}
}

@article{kreibich2004honeycomb,
  title={Honeycomb: creating intrusion detection signatures using honeypots},
  author={Kreibich, Christian and Crowcroft, Jon},
  journal={ACM SIGCOMM computer communication review},
  volume={34},
  number={1},
  pages={51--56},
  year={2004},
  publisher={ACM New York, NY, USA}
}

@inproceedings{lippmann2000evaluating,
  title={Evaluating intrusion detection systems: The 1998 DARPA off-line intrusion detection evaluation},
  author={Lippmann, Richard P},
  booktitle={Proceedings DARPA Information Survivability Conference and Exposition. DISCEX'00},
  volume={2},
  pages={12--26},
  year={2000},
  organization={IEEE}
}

@article{ullah2021design,
  title={Design and development of a deep learning-based model for anomaly detection in IoT networks},
  author={Ullah, Imtiaz and Mahmoud, Qusay H},
  journal={IEEE Access},
  volume={9},
  pages={103906--103926},
  year={2021},
  publisher={IEEE}
}

@inproceedings{hindy2020machine,
  title={Machine learning based IoT intrusion detection system: An MQTT case study (MQTT-IoT-IDS2020 dataset)},
  author={Hindy, Hanan and Bayne, Ethan and Bures, Miroslav and Atkinson, Robert and Tachtatzis, Christos and Bellekens, Xavier},
  booktitle={International networking conference},
  pages={73--84},
  year={2020},
  organization={Springer}
}

@inproceedings{ullah2020scheme,
  title={A scheme for generating a dataset for anomalous activity detection in iot networks},
  author={Ullah, Imtiaz and Mahmoud, Qusay H},
  booktitle={Canadian conference on artificial intelligence},
  pages={508--520},
  year={2020},
  organization={Springer}
}

@article{samarakoon20225g,
  title={5G-NIDD: A Comprehensive Network Intrusion Detection Dataset Generated over 5G Wireless Network},
  author={Samarakoon, Sehan and Siriwardhana, Yushan and Porambage, Pawani and Liyanage, Madhusanka and Chang, Sang-Yoon and Kim, Jinoh and Kim, Jonghyun and Ylianttila, Mika},
  journal={arXiv preprint arXiv:2212.01298},
  year={2022}
}

@inproceedings{moustafa2015unsw,
  title={UNSW-NB15: a comprehensive data set for network intrusion detection systems (UNSW-NB15 network data set)},
  author={Moustafa, Nour and Slay, Jill},
  booktitle={2015 military communications and information systems conference (MilCIS)},
  pages={1--6},
  year={2015},
  organization={IEEE}
}

@inproceedings{kao2015automatic,
  title={Automatic nids rule generating system for detecting http-like malware communication},
  author={Kao, Chia-Nan and Chang, Yung-Cheng and Huang, Nen-Fu and Liao, I-Ju and Liu, Rong-Tai and Hung, Hsien-Wei and Lin, Che-Wei},
  booktitle={2015 International Conference on Intelligent Information Hiding and Multimedia Signal Processing (IIH-MSP)},
  pages={199--202},
  year={2015},
  organization={IEEE}
}

@article{du2025harnessing,
  title={Harnessing Large Language Models for Automated Intrusion Detection Rule Generation in Cyber Range},
  author={Du, Lei and Li, Jiarui and Yan, Hao and Chai, Yuhan and Fang, Binxing and Gu, Zhaoquan},
  journal={IEEE Network},
  year={2025},
  publisher={IEEE}
}

@inproceedings{balasubramanian2024hex2sign,
  title={Hex2Sign: Automatic IDS Signature Generation from Hexadecimal Data using LLMs},
  author={Balasubramanian, Prasasthy and Ali, Tarek and Salmani, Mohammad and KhoshKholgh, Danial and Kostakos, Panos},
  booktitle={2024 IEEE International Conference on Big Data (BigData)},
  pages={4524--4532},
  year={2024},
  organization={IEEE}
}

@inproceedings{hu2024llm,
  title={A LLM-based agent for the automatic generation and generalization of IDS rules},
  author={Hu, Xiaowei and Chen, Haoning and Bao, Huaifeng and Wang, Wen and Liu, Feng and Zhou, Guoqiao and Yin, Peng},
  booktitle={2024 IEEE 23rd International Conference on Trust, Security and Privacy in Computing and Communications (TrustCom)},
  pages={1875--1880},
  year={2024},
  organization={IEEE}
}

\vfill

\end{document}